\documentclass[reprint,10pt,twocolumn,superscriptaddress,nofootinbib]{revtex4-1}
\usepackage[utf8]{inputenc}
\usepackage{graphicx}
\usepackage{bm}
\usepackage{amsmath,amssymb,amsbsy}
\usepackage{bbm,mathrsfs,yfonts}
\usepackage{float}
\usepackage{hyperref}
\hypersetup{
    colorlinks=true,
    linkcolor=blue,
    filecolor=magenta,      
    urlcolor=cyan,
}
\usepackage{silence}
\usepackage{mathtools}
\renewcommand{\ne}{\ensuremath{n_e}}
\newcommand{\Ni}{\ensuremath{N_i}}
\newcommand{\Nz}{\ensuremath{N}}

\newcommand{\dNz}{\ensuremath{\delta N}}

\newcommand{\neref}{\ensuremath{n_{e0}}}
\newcommand{\Nzref}{\ensuremath{N_{0}}}

\newcommand{\Pzp}{\ensuremath{P_{\perp}}}

\newcommand{\Tzp}{\ensuremath{T_{\perp}}}
\newcommand{\tep}{\ensuremath{T_{e\perp}}} 
\newcommand{\Tip}{\ensuremath{T_{i\perp}}}

\newcommand{\tepref}{\ensuremath{T_{e\perp0}}} 
\newcommand{\Tipref}{\ensuremath{T_{i\perp0}}} 

\newcommand{\Omegacz}{\ensuremath{\Omega}} 
 
\newcommand{\Omegaczref}{\ensuremath{\Omega_{0}}}

\renewcommand{\vec}[1]{{\boldsymbol{#1}}}
\newcommand{\bhat}{\hat{\vec b}}
\newcommand{\zhat}{\hat{\vec{e}}_z}

\newcommand{\nablap}{\ensuremath{\vec{\nabla}_\perp}}
\newcommand{\nablav}{\ensuremath{\vec{\nabla}}}

\newcommand{\ExB}{\ensuremath{\vec{E}\times\vec{B}} }
\newcommand{\gradB}{\ensuremath{\nablav B} }
\newcommand{\FullF}{\ensuremath{\text{Full-F }}}
\newcommand{\fullF}{\ensuremath{\text{full-F }}}

\newcommand{\dF}{\ensuremath{\delta\text{F}} }

\newcommand{\norm}[1]{{\|{#1}\|_{2}}}

\begin{document}
\title{Beyond the Oberbeck-Boussinesq and long wavelength approximation }
\author{M.\ Held}
\email[E-mail: ]{markus.held@uit.no}
\affiliation{Department of Mathematics and Statistics, UiT The Arctic University of Norway, 9037 Tromsø, Norway}
\affiliation{Department of Space, Earth and Environment, Chalmers University of 
Technology, SE-412 96 Gothenburg, Sweden}
\affiliation{Institute for Ion Physics and Applied Physics, 
                     Universit\"at Innsbruck, A-6020 Innsbruck, Austria}
\author{M.\ Wiesenberger}
\email[E-mail: ]{mattwi@fysik.dtu.dk}
\affiliation{Department of Physics, Technical University of Denmark, DK-2800 Kgs. Lyngby, Denmark}
\begin{abstract} 
We present the first simulations of a reduced magnetized plasma model that incorporates both arbitrary wavelength polarization and non-Oberbeck-Boussinesq effects. 
Significant influence of these two effects on the density, electric potential and \ExB vorticity and non-linear dynamics of interchange blobs are reported. 
Arbitrary wavelength polarization implicates so-called gyro-amplification that compared to a long wavelength approximation leads to highly amplified small-scale \ExB vorticity fluctuations. These strongly increase the coherence and lifetime of blobs and alter the motion of the blobs through a slower blob-disintegration.
Non-Oberbeck-Boussinesq effects incorporate plasma inertia, which substantially decreases the growth rate and linear acceleration of high amplitude blobs, while the maximum blob velocity is not affected.
Finally, we generalize and numerically verify unified scaling laws for blob velocity, acceleration and growth rate that include both ion temperature and arbitrary blob amplitude dependence. 
\end{abstract}
\maketitle
\section{Introduction}
The Oberbeck-Boussinesq and long wavelength approximation are two widely adopted simplifications in plasma theory with the aim to reduce the model's algebraic complexity and computational burden.

The Oberbeck–Boussinesq~\cite{oberbeck1879,boussinesq1903} approximation origins from neutral fluid dynamics and 
assumes that density variations only fully enter the buoyancy force and are everywhere else, i.e.  the inertia terms, assumed constant or linearly dependent on the temperature difference. This could be interpreted as if only a thin layer of a fluid is considered where no large inhomogeneities occur, resulting in the alternative term thin layer approximation.
\\
In reduced plasma fluid theories the Oberbeck-Boussinesq approximation enters through the polarization charge (or polarization current) that represents inertia. The Oberbeck-Boussinesq approximation is naturally inherent to \(\delta F\) models.
However, the validity of this assumption is questionable if large inhomogeneities appear. Such a large inhomogeneity occurs if large relative fluctuation amplitudes  emerge or if their size compares to the perpendicular e-folding length \(L_{\perp}\) so that \(k_\perp L_{\perp} \sim 1\).
\\
Large inhomogeneities can arise in a variety of magnetized plasma. They are typically observed in the edge and scrape-off layer of magnetically confined fusion plasmas~\cite{surko83,liewer85,ritz87,fonck93,endler95,mckee01,zweben07,zweben15,gao15,shao16,kobayashi16}.
\\
The physics of the non-Oberbeck-Boussinesq regime is rich, enabling radial zonal flow advection~\cite{held2018}, radially inhomogeneous zonal flows~\cite{held2019}, blob-hole symmetry breaking~\cite{kendl2015} and decreased blob or increased hole acceleration~\cite{wiesenberger2017}.

The long wavelength approximation~\cite{dubin1983} assumes that scales much larger than the gyro-radius \(k_\perp \rho \ll 1\) dominate the plasma dynamics. It typically neglects gyro-radius effects above powers \((k_\perp \rho)^2\).
This long wavelength ordering is inherent to reduced plasma descriptions, such as the drift-kinetic (or -fluid) description. However, it is also virtually always enforced for polarization effects in full-F gyro-kinetic or (-fluid) models, that are by construction \(k_\perp \rho \sim 1\). For full-F gyro-kinetic models this simplification is reasoned in the computational cost, while for full-F gyro-fluid models theory and closures have been only extended recently~\cite{held2020}.
\\
However, turbulence in magnetized plasmas, e.g. driven by an ion temperature gradient, is active at \(k_\perp \rho \sim 1\)~\cite{hennequin2004, tynan2009} and the interplay between small and large scales determines the overall plasma dynamics and transport. In particular, electron temperature gradient driven modes that are active on the much smaller electron gyro-radius interact with ion temperature gradient modes~\cite{howard2015}. Further,  order unity ion to electron temperature ratios appear~\cite{kocan2012,elmore2012,allan2016}, such that the ion gyro-radius is of the order of the drift scale. Thus \(k_\perp \rho\sim 1\) turbulence influences e.g. also drift-wave and interchange turbulence~\cite{kendl2018}.
This evidence suggests that the long wavelength approximation is inadmissible in turbulence models of magnetized plasmas.
\\
The application of the long wavelength approximation significantly reduces the lifetime and compactness of interchange blobs within the Oberbeck-Boussinesq regime~\cite{wiesenberger2014}. Further, the long wavelength approximation crucially affects the normal modes of ion temperature gradient or trapped electron mode instabilities~\cite{hammett92,dorland93,dominski17,mishchenko19}.

So far, the combined arbitrary wavelength and non-Oberbeck-Boussinesq regime is terra nova and it is unknown whether a synergy or superposition of known effects in the individual regimes exists.
In the following, we present the first numerical investigation of a reduced magnetized plasma model that captures both regimes consistently. To this end, we utilize a recently developed \fullF gyro-fluid model~\cite{held2020} that incorporates non-Oberbeck-Boussinesq and arbitrary wavelength polarization. Previous studies using gyro-fluid models were either limited to Oberbeck-Boussinesq~\cite{madsen2011} or long wavelength approximations~\cite{wiesenberger2014,kendl2015,held2016,held2018}. 
Gyro-kinetic studies relax the Oberbeck-Boussinesq approximation rarely consistently and then only in the long wavelength limit~\cite{Heikkinen08,Scott10}. Attempts to include also arbitrary wavelength effects are based on an unphysical ad-hoc Pad{\'e}-approximation that lacks a positive definite (e.g. quadratic) form of the kinetic energy~\cite{idomura03,bottino04,ku09,dominski17,mishchenko19}.
Studies based on drift-kinetic or -fluid models are inherently derived under the assumption $k_\perp \rho_i \ll 1$ (see for example \cite{madsen16,simakov03}) and are thus always in the long-wavelength limit. 
We limit our study to interchange blob dynamics since they are a well known and understood benchmark test case and the latter combined regimes are of particular relevance for filamentary transport in the scrape-off layer of magnetically confined fusion devices. By means of theoretical estimates and numerical experiments that spawn a large parameter space we show that arbitrary wavelength polarization overlays with non-Oberbeck-Boussinesq effects in their contribution to the nonlinear blob dynamics.  
On the one hand, arbitrary wavelength effects introduce gyro-amplification of small-scale \ExB vorticity. This manifests in strong small-scale \ExB shear flows at the blob edge that result in highly coherent and long-living blobs. 
On the other hand, non-Oberbeck-Boussinesq effects are accompanied by plasma inertia that influences the acceleration and growth rate of the blobs with increasing blob amplitude.

The remainder of the manuscript is organized as follows. The \fullF gyro-fluid model is presented in Section~\ref{sec:model}, where we elaborate on the accuracy of the arbitrary wavelength polarization closure and for comparison also introduce and discuss the long-wavelength and Oberbeck-Boussinesq approximated models. Further general relative error estimates for the Oberbeck-Boussinesq and long wavelength limit are derived that allow to estimate the validity of the latter approximations. On top of that, so-called gyro-amplification is introduced, which boosts  small-scale \ExB vorticity fluctuations when arbitrary wavelength polarization is retained. At the end of this section the invariants of our model are introduced and the blob initial condition is stated.
In Section~\ref{sec:scalinglaws} unified scaling laws for the blob center of mass velocity, acceleration and interchange growth rates are deduced from the full-F gyro-fluid model, its invariants and initial condition.
Our numerical experiments in Section~\ref{sec:numerical} reinforce the fundamental difference in the nonlinear dynamics of cold and hot blobs. We there also expand in detail on the influence of arbitrary wavelength and non-Oberbeck-Boussinesq effects on the blob dynamics, shape, pattern and compactness. Further, the previously derived unified scaling laws are verified by our numerical experiments.
Finally, we summarize our main findings in Section~\ref{sec:discussion}.

\section{Full-F gyro-fluid model}\label{sec:model}
We consider a \fullF gyro-fluid model, that is founded on the standard gyro-kinetic ordering~\cite{madsen2013} and fully encompasses effects down to the gyro-radius scale~\cite{held2020}. The gyro-fluid moment hierarchy is closed  by truncation, in particular an isothermal assumption. For the sake of simplicity, we neglect parallel dynamics and electromagnetic effects. 
The \fullF gyro-center continuity equations for a gyro-center species density 
\(\Nz\) \footnote{For convenience we omit the species subscript \(s\) and denote 
only explicit species with a subscript like electrons(e) or ions(i). Further, we use capital letters for gyro-center quantities and small letters for particle quantities, e.g. gyro-center density \(\Nz\) and particle density \(n\). 
} is given by
\begin{align}
\label{eq:continuityFF}
 \frac{\partial}{\partial t} \Nz  +\nablav \cdot \left[\Nz \left( \vec{U}_E + 
\vec{U}_{\nabla B} \right)\right] &=0,
\end{align}
where the gyro-center \ExB and \gradB drifts are
\begin{subequations}
\begin{align}
\label{eq:UExB}
  \vec{U}_E &:= \frac{\bhat \times \nablap (\psi_1 + \psi_2) }{B}, \\
\label{eq:UgradB}
   \vec{U}_{\nablav B} &:= \frac{\Tzp \bhat \times \nablap \ln (B/B_0)}{q B} ,
\end{align}
\end{subequations}
with the gyro-average and polarization part of the gyro-fluid potential 
\begin{subequations}
\begin{align}
\label{eq:psi1}
    \psi_1 &:=\Gamma_1 \phi, \\
\label{eq:psi2}
    \psi_2 &:=-\frac{q}{2 m \Omegacz^2}  |\nablap \sqrt{\Gamma_0}\phi|^2 .
\end{align} 
\end{subequations}
The magnetic field points into 
z and its magnitude is varying in x-direction \( B(x) := B_0 \left[(x-x_0)/R_0 
+1 \right]^{-1}\). Here we defined the reference radius  \(R_0\), the initial 
position \(x_0\)  and the reference magnetic field magnitude \(B_0\). The magnetic field unit vector is defined by \(\bhat:=\vec{B}/B = \zhat\). It is utilized for the perpendicular projection of a vector \(\vec{h}\) according to \(\vec{h}_\perp :=-\bhat \times (\bhat \times \vec{h}) = \mathbb{P} \cdot \vec{h}\), where \(\mathbb{P} := \vec{g} - \bhat \bhat\) and  \(\vec{g}\) is the projection and metric tensor, respectively. We also introduced the particle charge  \(q\), mass \(m\), gyro-frequency \(\Omegacz:=q B/m\) and perpendicular gyro-center temperature  \(\Tzp\).
\\
The finite Larmor radius (FLR) and polarization operators are included through Pad{\'e} 
approximations~\cite{held2020}
\begin{subequations}
\begin{align}
\label{eq:gamma1}
 \Gamma_1 &:=(1-\rho^2/2 \Delta_\perp)^{-1}, \\
 \label{eq:gamma0}
 \Gamma_0 &:=(1-\rho^2 \Delta_\perp)^{-1},
\end{align}
\end{subequations}
and are taken in this work in the \(\rho =const.\) limit. As a consequence these operators are self-adjoint. Here, we introduce the thermal gyro-radius and also define the the drift scale
\begin{align} \label{eq:gyro-radius}
     \rho &:= \frac{\sqrt{\Tzp m}}{q B}, \\ \label{eq:drift-scale} 
 \rho_{s0} &:= \frac{\sqrt{T_e m_i}}{eB},
 \end{align}
 respectively. The utilized Pad{\'e} 
approximations in the  FLR and polarization operators (Eq.~\eqref{eq:gamma1} and~\eqref{eq:gamma0}) are excellent approximations to the exact operators 
\begin{align}
\label{eq:gamma1ex}
    \Gamma_1^{ex} &:= \exp{(\rho^2/2 \Delta_\perp)}, \\
\label{eq:gamma0ex}
    \Gamma_0^{ex} &:= \exp{(\rho^2\Delta_\perp)I_0(\rho^2 \Delta_\perp)},
\end{align}
where \(I_0\) is the zeroth order modified Bessel function~\cite{held2020}. In particular, the Pad{\'e} approximations are fully accurate up to \(\mathcal{O}(k_\perp^2 \rho^2)\) and closely mimic the behaviour at arbitrary wavelengths~\cite{held2020}.
\\
We emphasize that in the gyro-center continuity Eq.~\eqref{eq:continuityFF} only the compression of the \ExB drift, \(\nablav \cdot 
\vec{U}_E= - \frac{1}{B_0 R_0}\partial_y (\psi_1 + \psi_2)\),  is non-vanishing. This is important for the blob dynamics at small fluctuation amplitudes~\cite{kube2016,wiesenberger2017}.
\\
The \fullF gyro-fluid model is completed by the Poisson equation
\begin{align}
\label{eq:PoissonFF}
\sum_s q \Nz
- \nablav \cdot \left(\vec{P}_1 + \vec{P}_2\right) &=   0,
\end{align}
with the polarization densities~\cite{held2020}
\begin{subequations}
\begin{align}
\label{eq:P1FF}
\vec{P}_1  &=  - \sum_s q   \nablap  \Gamma_1 \rho^2 \Nz /2,\\
\label{eq:P2FF}
 \vec{P}_2 &=- \sum_s   \left(\sqrt{\Gamma_{0}} \frac{q\Nz}{\Omega B} 
\sqrt{\Gamma_{0}}\nablap \phi\right).
\end{align}
\end{subequations}
Note that the use of Pad{\'e} approximations allows us to write the 
polarization charges as simple divergence relations of the polarization 
densities (e.g. 
\(-\nablav \cdot \vec{P}_1 =\sum_s q  (\Gamma_1 -1) N \equiv \nablav \cdot \sum_s 
q  \nablap \Gamma_1\rho^2 N/2\)).
\\
In Eq.~\eqref{eq:P2FF} and~\eqref{eq:UExB} we adopt the second order accurate Pad{\'e} approximation of Ref.~\cite{held2020} that gives rise to the square root of the  Pad{\'e} approximated polarization operator \(\sqrt{\Gamma_0}\). A square root operation  could be avoided by using the fourth  order accurate Pad{\'e} approximation of Ref.~\cite{held2020}. However, this  results in a fourth order elliptic equation instead of a second order elliptic equation in Eq.~\ref{eq:PoissonFF} (cf. Ref.~\cite{held2020}). Further, it does not resemble the commonly used second order accurate Pad{\'e} approximation of Eq.~\eqref{eq:gamma0} of Oberbeck-Boussinesq approximated gyro-fluid models in the Oberbeck-Boussinesq limit.
\\
The herein utilized gyro-fluid model resorts to a constant thermal gyro-radius (\(\rho=const.\)) and a magnetic field aligned coordinate system (Cartesian coordinates and \(\bhat=\zhat\)). These simplifications ease the analytical and numerical treatment of the arbitrary wavelength polarization closure. 
The \(\rho=const.\) approximation in the FLR and polarization operators (Eqs.~\eqref{eq:gamma1} and ~\eqref{eq:gamma0}) permits to commute the gyro-radius with the perpendicular Laplacian, so that these operators are self-adjoint.
The additional choice of a magnetic field aligned coordinate system allows us to commute the polarization operators through the divergence \(\vec{\nabla}\cdot \sqrt{\Gamma_0} \vec{f} = \sqrt{\Gamma_0}\vec{\nabla}\cdot \vec{f}\) and spatial derivatives \( \sqrt{\Gamma_0} \vec{\nabla}_\perp f = \vec{\nabla}_\perp \sqrt{\Gamma_0} f \). In general this is permitted due a spatially dependence in the thermal gyro-radius \(\rho\) and in the generally non-orthogonal projection tensor \(\mathbb{P}\).
However, we remark that cylindrical coordinates  and the straight field line approach are commonly exploited for realistic three dimensional computations with toroidal magnetic fields, such as Tokamaks. For this approach the simplifications above can be readily exploited within the \(\rho=const.\) approximation.
\\
We emphasize that the arbitrary wavelength polarization closure is by construction only fully accurate up to \(\mathcal{O}(k_\perp^2 \rho^2)\) but resembles the Pad{\'e} approximated Oberbeck-Boussinesq limit (Eq.~\eqref{eq:gamma0}) at arbitrary \(k_\perp \rho\). This enables us to bridge the gap between long wavelength approximated \fullF and arbitrary wavelength Oberbeck-Boussinesq approximated \fullF (or \(\delta F\))  gyro-fluid models. 
The numerical implementation and study of a polarization closure that is fully accurate or achieves better accuracy to arbitrary \(k_\perp \rho\)  is shifted to future work. In Sec. \ref{sec:polacc} we elaborate more on the accuracy of the chosen arbitrary wavelength polarization closure.
\\
We remark that we do not consider the ad-hoc approximation
\(\vec{\nabla} \cdot \vec{P_2}\approx -\sum_s \frac{1}{1- \vec{\nabla} \cdot (\rho^2 \nablap)}\vec{\nabla}\cdot (\frac{q \Nz}{\Omegacz B}\nablap \phi)\) ~\cite{idomura03,bottino04,ku09,dominski17,mishchenko19} since it can not be derived from a particular choice of \(\psi_2 \) by means of field theory.
\subsection{Accuracy of the arbitrary wavelength polarization closure}\label{sec:polacc}
\subsubsection{Exact arbitrary wavelength polarization}
\begin{figure*}[ht]
\centering
\includegraphics[width=0.8\textwidth]{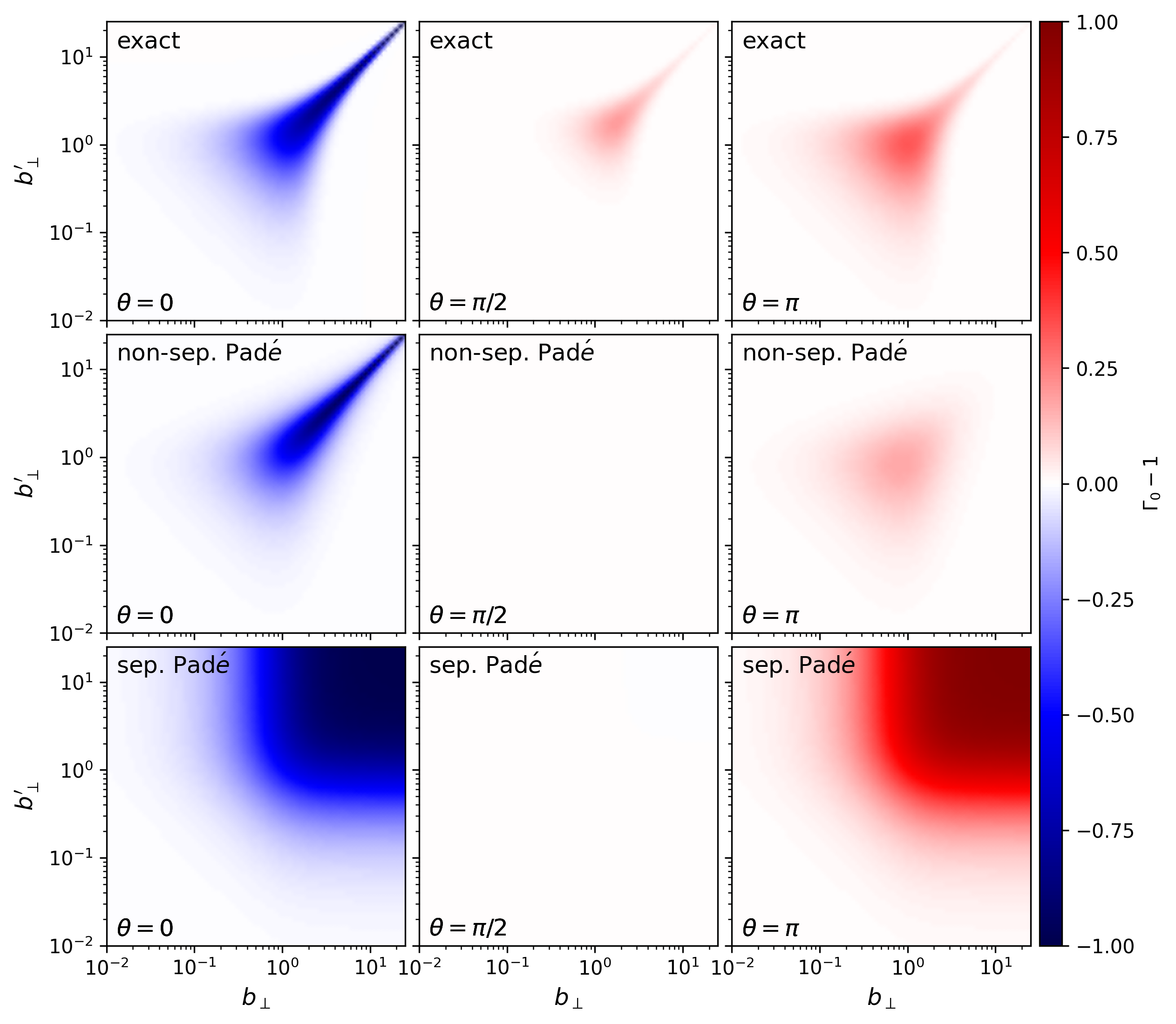}
\caption{The exact, non-separable and separable Pad{\'e} approximated polarization kernels (Eqs.~\eqref{eq:gamma0fullFexac},~\eqref{eq:gamma0fullFPadenonsep} and~\eqref{eq:gamma0fullFPadesep}, respectively) are shown for three different phase angles \(\theta=(0,\pi/2,\pi)\). In contrast to the exact polarization kernel its Pad{\'e} approximations vanish for orthogonal wave-vectors \(\theta=\pi/2\). Further the separable Pad{\'e} approximation is not able to capture the sharpening of the finite tail of the exact and non-separable Pad{\'e} approximation towards arbitrary wavelengths for parallel (\(\theta=0\)) and anti-parallel (\(\theta=\pi\)) wave-vectors. }
\label{fig:polkernel}
\end{figure*}

The exact expressions for the arbitrary wavelength polarization treatment with a near Maxwellian gyro-center distribution function have been derived recently in Ref.~\cite{held2020}, which provide the gyro-moment expressions in Fourier space as well as its Taylor series expansion in configuration space. In the following we list the arbitrary wavelength expressions of the polarization part of the basic gyro-fluid potential
\begin{align}
\label{eq:psi2exac}
  \psi_2  &= \frac{q }{ 2\rho^2 m \Omega^2} \sum_{\vec{k},\vec{k}'} \left[\Gamma_0^{ex}(\rho \vec{k},-\rho \vec{k}')-1\right] \phi_{\vec{k}} \phi_{\vec{k}'} e^{\mathrm{i} \vec{K}\cdot \vec{X}}, 
\end{align}
and its associated polarization charge density
\begin{align}
\label{eq:polchargeex}
 - \vec{\nabla}\cdot \vec{P}_2 &=
   \sum_s\frac{q^2}{m \rho^2 } \sum_{\vec{k},\vec{k}'}  
  \left[\Gamma_0^{ex}(\rho\vec{K},\rho\vec{k})-1\right]\left(\frac{N}{\Omega^2}\right)_{\hspace{-1mm}\vec{k}'} \hspace{-2mm}\phi_{\vec{k}}e^{\mathrm{i} \vec{K}\cdot\vec{X}} ,
\end{align}
which results from variational calculus~\cite{held2020}.  
Without loss of generality in the following discussion we adhere to \(\rho=const.\).
Here, we defined the wave-vector \(\vec{K}:=\vec{k}+\vec{k}'\) and the non-Oberbeck-Boussinesq form of the exact polarization operator
\begin{align}
\label{eq:gamma0fullFexac}
     \Gamma_0^{ex}(\vec{b},\vec{b}')&:=1-e^{\frac{-(b_\perp^2 + b_\perp'^2)}{2}} \left[e^{\vec{b}_\perp\cdot \vec{b'}_\perp}  - I_0 ( b_\perp  b_\perp')\right].
\end{align}
First, we notice that  Eqs.~\eqref{eq:gamma0fullFexac}, i.e. \(\Gamma_0^{ex}(\vec{b},\vec{b}') -1\), is a convolution kernel of infinite rank due to the bracket \(e^{\vec{b}_\perp\cdot \vec{b'}_\perp}  - I_0 ( b_\perp  b_\perp')\) and thus is not separable. 
Second, solving the exact polarization part of the basic gyro-fluid potential (Eq.~\eqref{eq:psi2exac}) and the gyro-fluid Poisson Eq.~\eqref{eq:polchargeex} necessitates an efficient numerical solution of a convolution and deconvolution, respectively.
By contrast a separable approximation 
to the exact convolution kernel allows to bypass these cumbersome convolutions within a treatment in configuration space.
\\
In Fig.~\ref{fig:polkernel} (top row) we show the exact polarization kernel for three meaningful phase angles \(\theta\). Here, the phase angle \(\theta\) is the angle between \(\vec{b}_\perp\) and \(\vec{b}_\perp'\), so that \(\vec{b}_\perp\cdot \vec{b}_\perp' = b_\perp b_\perp' \cos{(\theta)}\) holds. 
The exact polarization kernel features a sharp finite tail for all presented phase angles around roughly \(b_\perp \sim b_\perp'\). This sharp tail appears broadly around \(b\gtrsim 1\) and extends narrowly to \(b\rightarrow \infty\).
\\
The polarization operator of Eq.~\eqref{eq:gamma0fullFexac} reduces to the conventional and exact polarization operator  (Eq.~\eqref{eq:gamma0ex}) in the Oberbeck-Boussinesq limit. 
There, Eq.~\eqref{eq:psi2exac} vanishes and we neglect the spatial dependence in the gyro-fluid moment variables and the magnetic field magnitude  in the polarization charge density, so that  $\Gamma_0^{ex}(\vec{b},\vec{b}')\approx \Gamma_0^{ex}(\vec{b},\vec{b})= \Gamma_0^{ex} (b)$.
The long wavelength limit expressions are derived by Taylor expanding the exact polarization kernel  to \(\mathcal{O}(\vec{b}_\perp,\vec{b}_\perp')\) so that \(
\Gamma_0^{ex}(\vec{b},\vec{b}')-1 \approx - \vec{b}_\perp \cdot \vec{b}_\perp'\).
\subsubsection{Non-separable Pad{\'e} approximation}
Exploiting a symmetric bivariate  Pad{\'e} approximation of order \((1,2)\) to each term of the exact operator (by expanding $\Gamma_0^{ex}(t b_\perp,t b_\perp',\theta) $ around $t=0$ and setting $t=1$ afterwards) yields
\begin{align}
\label{eq:gamma0fullFPadenonsep}
\Gamma_0^{ex}(\vec{b},\vec{b}')-1&\approx
-\frac{\vec{b}_\perp \cdot \vec{b}_\perp'}{\left(1+ \frac{b_\perp^2 + b_\perp'^2}{2}\right)\left(1+\frac{b_\perp^2 + b_\perp'^2}{2}- \vec{b}_\perp \cdot \vec{b}_\perp'\right)}.
\end{align}
Note that the bivariate Pad{\'e} approximation is not separable, fully accurate to   \(\mathcal{O}(k_\perp^{2} \rho^{2})\) and resembles Eq. ~\eqref{eq:gamma0} in the Oberbeck-Boussinesq limit. In Fig.~\ref{fig:polkernel} (center row) it is shown that the sharp tail of the exact polarization kernel is well approximated to arbitrary wavelengths for parallel and anti-parallel wave-vectors (\(\theta=0\) and \(\theta=\pi\)). However for orthogonal wave-vectors the kernel vanishes and does not resemble the finite peak of the exact polarization kernel.
 \\
Note that a bivariate Pad{\'e} approximation of order \((1,2)\) to the full expression yields \(\Gamma_0^{ex} (\vec{b},\vec{b}')-1 \approx -\frac{\vec{b}_\perp \cdot \vec{b}_\perp'}{1+\vec{b}_\perp \cdot \vec{b}_\perp'}\), which is not separable and also less accurate. Further, a nested Pad{\'e} approximation of order \((1,2)\) also produces a non-separable approximation, which is too complex to implement while being less accurate than Eq.~\eqref{eq:gamma0fullFPadenonsep}.
\subsubsection{Separable Pad{\'e} approximation}
The Pad{\'e} approximation proposed in Ref.~\cite{held2020} is by construction separable, exactly accurate to order \(\mathcal{O}(k_\perp^{2n} \rho^{2n})\) and reduces in the Oberbeck-Boussinesq limit to the Pad{\'e} approximated polarization operator.
The most simple  yet \(\mathcal{O}(k_\perp^{2} \rho^{2})\) accurate form of the Pad{\'e}  approximation is~\cite{held2020}
\begin{align}
\label{eq:gamma0fullFPadesep}
\Gamma_0^{ex}(\vec{b},\vec{b}')-1&\approx -\frac{\vec{b}_\perp \cdot \vec{b}_\perp'}{\sqrt{1+ b_\perp^2 }\sqrt{1+ b_\perp'^2}} .
\end{align} 
Observe that the latter separable operator resembles the long wavelength limit and the Oberbeck-Boussinesq limit of \(\Gamma_0\). Further, we find the polarization relevant terms in configuration space (Eqs.~\eqref{eq:psi2} and~\eqref{eq:P2FF} with Eq.~\eqref{eq:gamma0}) from an inverse Fourier transform of Eqs.~\eqref{eq:psi2exac} and~\eqref{eq:polchargeex} with Eq.~\eqref{eq:gamma0fullFPadesep}.
\\
The separable Pad{\'e} approximated polarization kernel is depicted in Fig.~\ref{fig:polkernel} (bottom row). We notice that the sharp region of the exact polarization kernel, i.e for \(b_\perp \sim b_\perp \gg 1\), is not well captured by the separable Pad{\'e} approximation. Similarly to the non-separable  Pad{\'e} approximated polarization kernel its separable equivalent vanishes for orthogonal wave-vectors.
\
\subsection{Oberbeck-Boussinesq limit}
The Oberbeck-Boussinesq approximated \fullF gyro-fluid model arises by assuming (i) small relative fluctuation amplitudes in the gyro-center density and perpendicular pressure, \(|\delta \Nz|\ll 1\) and \(|\delta\Pzp|\ll 1\), (ii) large density and perpendicular pressure gradient lengths, \(k_\perp \Nz/|\vec{\nabla} \Nz| \gg 1\) and \(k_\perp \Pzp/|\vec{\nabla} \Pzp| \gg 1\), (iii)  small spatial variations in the magnetic field magnitude\footnote{so that e.g. \(\vec{U}_E \cdot \vec{\nabla} \Nz \approx \Nzref B_0^{-1} \bhat \times \nablap \Gamma_1 \phi  \cdot \vec{\nabla} \dNz\)} and (iv) a vanishing second order polarization term  (\(\psi_2 = 0\)) in the gyro-fluid \ExB drift of Eq.~\eqref{eq:UExB} to restore energetic consistency. 
Here, we introduced the relative fluctuation \(\delta Q := Q/Q_0-1\) of a quantity \(Q\) with respect to its stationary part \(Q_0\).
These assumptions result in a  Oberbeck-Boussinesq approximated \fullF gyro-fluid model that is similar to the \dF gyro-fluid model. We emphasize that the Oberbeck-Boussinesq approximation necessitates the splitting of the gyro-center distribution function into a stationary and relative fluctuation part~\cite{held2020}. This stationary part of the gyro-center distribution function then appears as a gyro-moment quantity in the polarization density.  In general this stationary state is not known a priori or there is no stationary state at all. 
For the first case the stationary state must be determined by a (non-Oberbeck-Boussinesq or \fullF) model that avoids such a separation. For the second case an approximate stationary state must be assumed.
\\
The Oberbeck-Boussinesq approximated \fullF gyro-fluid model consists of the gyro-center continuity equations
\begin{align}
\label{eq:continuitydF}
 \frac{\partial}{\partial t} \Nz  + \Nzref \nablav\cdot\vec{U}_E + \left( \vec{U}_E + 
\vec{U}_{\nabla B} \right)\cdot\nablav\Nz &=0,
\end{align}
and the \fullF  Poisson equation of Eq.~\eqref{eq:PoissonFF} with the approximated 2nd order polarization density and polarization part of the gyro-fluid potential (resulting from Eq.~\eqref{eq:P2FF} and~\eqref{eq:psi2})~\cite{held2020}
\begin{align}
\label{eq:P2psi2OB}
 \vec{P}_2 &\approx- \sum_s \frac{q \Nzref}{\Omegaczref B_0}  \Gamma_{0} \nablap \phi,
 &
 \psi_2 &\approx 0.
\end{align}
Note that we do not drop the \ExB compression term in Eq.~\eqref{eq:continuitydF}, in contrast to the original Oberbeck-Boussinesq approximation.
\subsection{Long wavelength limit}
In the long wavelength limit we assume 
\(\rho^2 \Delta_\perp \Omega_E \sim \Omega_E\), but also
\(\rho^2 B_0 \nablap \ln p_\perp \cdot \nablap \Omega_E  \sim \nablap \ln n \cdot \nablap \phi \) in 
the polarization terms of Eq.~\eqref{eq:P2FF} and~\eqref{eq:psi2}, which yields
\begin{align}
\label{eq:P2psi2LWL}
  \vec{P}_2  &\approx  -\frac{q \Nz}{\Omega B} \nablap \phi,  
  &
  \psi_2 &\approx -\frac{q}{2 m \Omegacz^2}  |\nablap \phi|^2 .
\end{align}
We introduced the \ExB vorticity 
\begin{align}
    \Omega_E&:= \bhat \cdot \nablav \times \vec{u}_E,
\end{align}
where $\vec u_E = (\bhat \times \nablav \phi)/B$. 
The long wavelength approximation of Eq.~\eqref{eq:P2psi2LWL} is in line with previous full-F gyro-fluid studies that equally adopt the long wavelength limit only in the 2nd order polarization 
terms~\cite{wiesenberger2014,kendl2015,held2016, kendl2017, held2018, held2019}. By contrast, a full-F drift-fluid formalism is recovered if the long wavelength approximation is additionally applied to the 1st order polarization (also known as diamagnetic) terms of Eq.~\eqref{eq:P1FF} and~\eqref{eq:psi1} together with replacing the gyro-center density \(\Nz\) by a particle density \(n\) in Eq.~\eqref{eq:P2psi2LWL}.

\subsection{Long wavelength and Oberbeck-Boussinesq limit}
For comparison also the long wavelength limit of the Oberbeck-Boussinesq approximated \fullF gyro-fluid model is considered, where the polarization terms of Eq.~\eqref{eq:P2FF} and~\eqref{eq:psi2} are approximated according \begin{align}
\label{eq:P2psi2OBLWL}
    \vec{P}_2  &\approx   -\frac{q \Nzref}{\Omegaczref B_0} 
    \nablap \phi,
    &
    \psi_2 &\approx0.
\end{align}  
\\
For the sake of clarity, we summarize the various treatments of the polarization terms that appear in the original \fullF gyro-fluid model, as well as its three approximations in Table~\ref{table:poltreatment}.
\begin{table}
\caption{Summary of polarization treatment of the original \fullF gyro-fluid models as well as its long wavelength (LWL) and Oberbeck-Boussinesq (OB) approximations}
\begin{ruledtabular}
  \label{table:poltreatment}
 \begin{tabular}{l | l | l  }   
  GF model&  \(\vec{P_2}\) &\(\psi_2\) \\
  \hline
 \FullF & \(  -\sqrt{\Gamma_0} \frac{q \Nz}{\Omegacz B} \sqrt{\Gamma_0}\nablap 
\phi\) & \(-\frac{q}{2 m \Omegacz^2}  |\nablap \sqrt{\Gamma_0}\phi|^2 \) \\
 \FullF + OB &  \( -\frac{q \Nzref}{\Omegaczref B_0} \Gamma_0 \nablap \phi\)  & 
\(0\)\\
\FullF + LWL & \(  -\frac{q \Nz}{\Omegacz B} \nablap \phi\) & \(-\frac{q}{2 m 
\Omegacz^2}  |\nablap\phi|^2 \)\\
 
 \FullF + OB + LWL  &  \( -\frac{q \Nzref}{\Omegaczref B_0} \nablap \phi\) &  
\(0\)\\
\end{tabular}
 \end{ruledtabular}
\end{table}
\subsection{Gyro-amplification} \label{sec:amplification}
\begin{figure}[ht]
\centering
\includegraphics[width=0.475\textwidth]{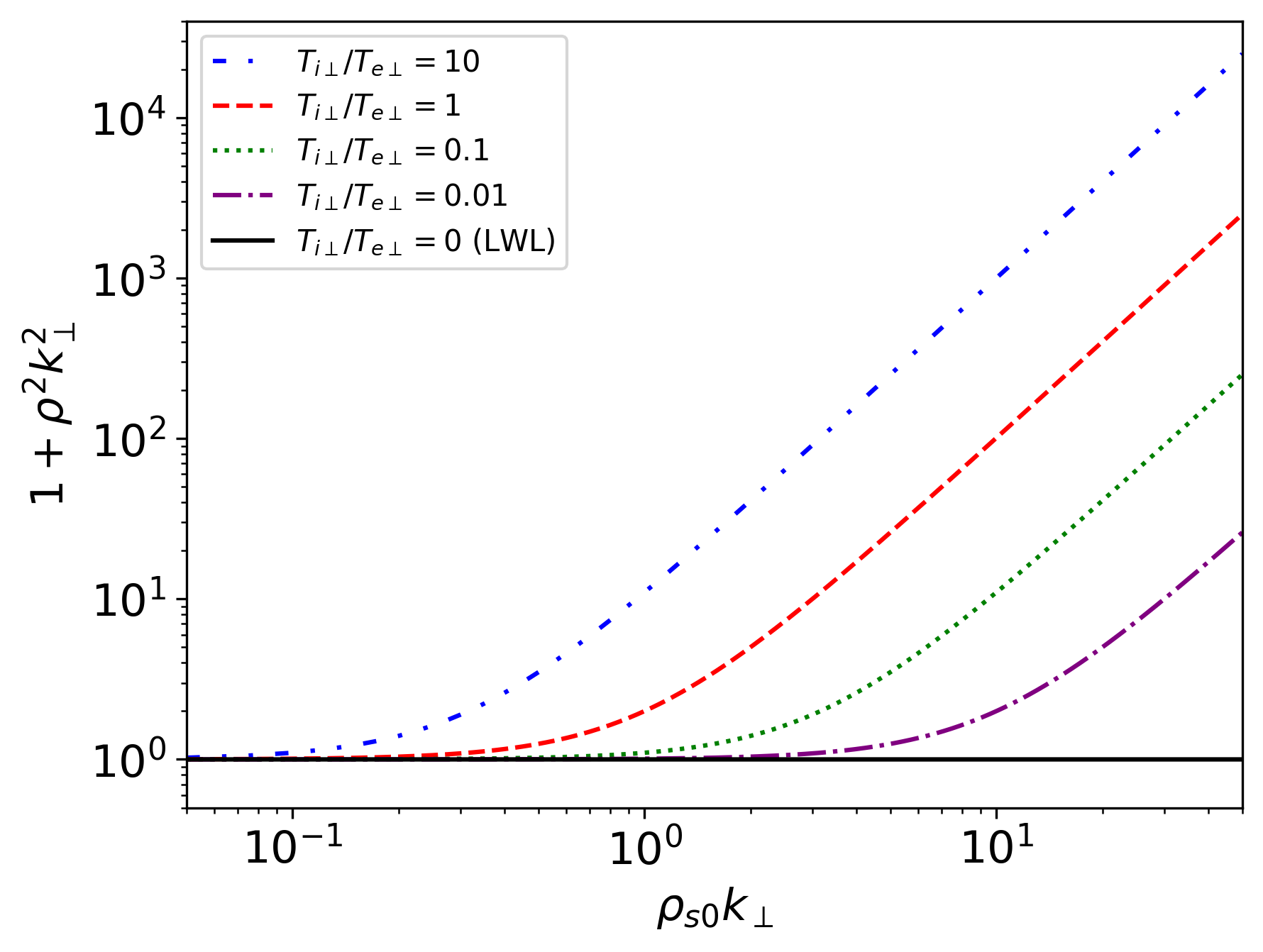}
\caption{The arbitrary wavelength polarization factor \(1+\rho^2 k_\perp\) arising in the Poisson equation  is shown for different values of \(T_{i\perp}/T_{e\perp}\). The inclusion of arbitrary perpendicular wavelength effects leads to a enhancement of the \ExB vorticity since \(\Omega_{E,k}\sim (1+\rho^2 k_\perp^2)\)  }
\label{fig:gyroamp}
\end{figure}

Gyro-amplification refers to the increase of \ExB vorticity when retaining arbitrary perpendicular wavelength polarization. This can be understood from solving the Oberbeck-Boussinesq approximated Poisson equation with Eq.~\eqref{eq:P2psi2OB} for the \ExB vorticity \( \Omega_E\sim \Delta_\perp \phi \).
This results in an additional factor \(\Omega_{E,k}\sim (1+\rho^2 k_\perp^2)\) for arbitrary wavelength polarization, stemming from the use of the polarization operator $\Gamma_0$~\eqref{eq:gamma0}.
 Thus, \ExB vorticity structures at and below the drift scale $\rho_{s0}$
 are greatly enhanced for typical values of \(T_{i\perp}/T_{e\perp}=1\), which is depicted in Fig.~\ref{fig:gyroamp}. We will study this behaviour further in Section~\ref{sec:numerical}, where we will find that the
 amplification is typically around a factor $10$ at the dominant scale in comparison to the long wavelength treatment.
\subsection{Relative error estimates for the Oberbeck-Boussinesq and long wavelength approximation}
In order to quantify the applicability of the long wavelength and Oberbeck-Boussinesq approximation rigorously we introduce relative errors for the respective regime. 
These relative errors are deduced from the particular approximation in the Poisson Eq.~\eqref{eq:PoissonFF} by expanding up to \(\mathcal{O}(\rho^2 k_\perp^2 )\). Additionally, spatial variations in the magnetic field magnitude are assumed to be small.
\\
According to this principle, the relative errors of the long wavelength and Oberbeck-Boussinesq approximation are derived to
\begin{widetext}
\begin{subequations}
\begin{align}
\label{eq:epsLWL}
     \epsilon_{LWL} &:= \frac{\norm{ \rho^2 B_0  \left(\nablap \ln p_\perp \cdot \nablap \Omega_E  +  \Delta_\perp\Omega_E\right)}}{\norm{ \nablap \ln n \cdot \nablap \phi  +   B_0 \Omega_E + B_0 \rho^2\left( \nablap \ln p_\perp \cdot \nablap \Omega_E + \ \Delta_\perp \Omega_E\right)}},
\\
\label{eq:epsOB}
    \epsilon_{OB} &:= \frac{\norm{\nablap \ln n \cdot \nablap \phi  +   \frac{\delta n}{1+ \delta n} B_0\Omega_E + B_0 \rho^2\left( \nablap \ln p_\perp \cdot \nablap \Omega_E + \frac{\delta p_\perp}{1+\delta p_\perp } \Delta_\perp \Omega_E\right)}}{\norm{\nablap \ln n \cdot \nablap \phi  +  B_0  \Omega_E + B_0 \rho^2\left( \nablap \ln p_\perp \cdot \nablap \Omega_E + \ \Delta_\perp \Omega_E\right)}},
 \end{align}
 \end{subequations}
\end{widetext}
where \(\norm{f}\) denotes the 2-norm.
The relative error \(\epsilon_{LWL} \) depends on the inverse perpendicular temperature gradient  length, the thermal gyro-radius and the spatial scale of the electric potential (and thus the \ExB vorticity). 
The relative error \(\epsilon_{LWL}\) is large if \(\rho^2 \Delta_\perp \Omega_E \sim \Omega_E\) or if  
\(\rho^2 B_0 \nablap \ln p_\perp \cdot \nablap \Omega_E  \sim \nablap \ln n \cdot \nablap \phi \). In the Oberbeck-Boussinesq limit \(\epsilon_{LWL}\) of Eq.~\eqref{eq:epsLWL} recovers \(\epsilon_{LWL|OB}\) of Eq.~\eqref{eq:epsLWLOB}.
\\
An error of order unity is approached if one of the following conditions is fulfilled: \(\delta n  \sim 1\),  \(\delta p_\perp  \sim 1\), \(\nablap \ln n \cdot \nablap \phi  \sim B_0  \Omega_E\) or \(\nablap \ln p_\perp \cdot \nablap \Omega_E \sim \Delta_\perp  \Omega_E\). 
 In the long wavelength limit \(\epsilon_{OB}\) of Eq.~\eqref{eq:epsOB} recovers \(\epsilon_{OB|LWL}\) of Eq.~\eqref{eq:epsOBLWL}.

Analogously, we define the relative error for long wavelength approximation under the condition of the Oberbeck-Boussinesq approximation and vice versa by
\begin{subequations}
\begin{align}
\label{eq:epsLWLOB}
 \epsilon_{LWL |OB}&:= \frac{\norm{\rho^2 \Delta_\perp \Omega_E}}{\norm{\Omega_E+\rho^2 \Delta_\perp \Omega_E}} ,
 \\
 \label{eq:epsOBLWL}
  \epsilon_{OB |LWL} &:= \frac{\norm{\nablap \ln n \cdot \nablap \phi  +  \frac{\delta n}{1+ \delta n} B_0 \Omega_E}}{\norm{ \nablap \ln n \cdot \nablap \phi  +   B_0 \Omega_E}} .
\end{align}
\end{subequations}
Note that \(\epsilon_{LWL |OB} \) depends on the thermal gyro-radius and the spatial scale of the electric potential (and consequently the \ExB vorticity). More specifically, the long wavelength approximation is questionable if \(\rho^2 \Delta_\perp \Omega_E \sim \Omega_E\).
\\
The relative error \(\epsilon_{OB |LWL} \) depends on both the inverse density gradient length and the relative density fluctuation amplitude.
 The Oberbeck-Boussinesq approximation is questionable if \(\delta n \sim 1\)~\footnote{More specifically, \(|\delta n/(1+\delta n)| \sim 1/2\) so that e.g. for blobs and holes \(\delta n \sim 1\) and \(\delta n \sim -1/3\), respectively. }  or \(\nablap \ln n \cdot \nablap \phi  \sim B_0  \Omega_E\).

\subsection{Conserved quantities}
In the absence of dissipation and surface integral terms the presented gyro-fluid models possess a number of conserved quantities. These are 
the gyro-center total particle number $M$, total polarization charge $Q$, Helmholtz free energy $F$ and  mechanical energy \(Z\)
\begin{subequations}
\begin{align}
\label{eq:M}
 M(t) &:= \int dA (\Nz - \Nzref),\\
 \label{eq:Q}
 Q(t) &:= \sum_s q  M(t) ,\\
 \label{eq:F}
 F(t) &:= \sum_s \left[E_k (t) - \Tzp  S(t) \right], \\
 \label{eq:Z}
 Z(t) &:= \sum_s \left[ E_k (t) +  H(t) \right] ,
\end{align}
\end{subequations}
where here and in the following we integrate over the entire domain.
Here, we defined the kinetic energy $E_k$, the entropy $S$ and the potential energy $H$
\begin{subequations}
\begin{align}
\label{eq:Ekin}
  E_k(t) &:= -\int dA q N \psi_2,\\
  \label{eq:Entropy}
  S(t) &:= -\int dA \left[N \ln (N/\Nzref) - (N-\Nzref)  \right],\\
\label{eq:Epot}
  H(t) &:= -\Tzp \int dA (\Nz - \Nzref) \ln (B_0/B).
\end{align}
\end{subequations}
The kinetic energy and the entropy obey \(E_k(t) \geq0\) and \(S(t) \leq0\), respectively. 
Note that in the Oberbeck-Boussinesq limit Eqs.~\eqref{eq:Entropy} and ~\eqref{eq:Epot} reduce to
\begin{subequations}
\begin{align}
S(t) &\approx -\int dA  (N-\Nzref)^2/2, \\
H(t) &\approx - \Tzp \int dA (\Nz - \Nzref) (x-x_0)/R_0,
\end{align}
\end{subequations}
by Taylor expanding for small fluctuation amplitudes or small magnetic field variations, respectively. 
\subsection{Initialization}
We will consider two initial conditions that differ in the initial ion gyro-center density \(\Ni(\vec{x},0)\) and as a consequence the initial electric potential \(\phi(\vec{x},0)\). However,
the electron density field \(\ne(\vec{x},0)\) is always initialized by a Gaussian 
\begin{align}
\label{eq:init_ne}
 \ne(\vec{x},0) &= \neref + \Delta \ne \exp\left[ -\frac{(\vec{x}-\vec{x}_0)^2 
}{2\sigma^2}\right] ,
\end{align}
with size \(\sigma\), amplitude \(\Delta \ne\) and initial position 
\(\vec{x}_0\) on top of a constant background \(\neref\).
\\
The initial electron density of Eq.~\eqref{eq:init_ne} determines the initial total particle number \(M_e(0) = 2 \pi 
\sigma^2 \Delta \ne\) and the initial electron entropy  \(S_e(0) = - 2 \pi \sigma^2 f(\Delta \ne)\). 
Here, we defined the function \(f(x):= -2x+(1+x)\ln(1+x) - \text{Li}_2(-x)\) where  \(\text{Li}_n(x)\) is the polylogarithm function. For small relative fluctuation amplitudes the initial electron entropy recovers its \(\delta F\) limit \( S_e(0)\approx -\pi/2 \sigma^2 \Delta \ne^2\).
The initial electron potential energies is \(H_e(0) \approx0\), where we 
used again the Taylor expanded integral.
\subsubsection{Non-rotating Gaussian}
For the first initial condition the initial electric field vanishes  (\(\nablap\phi(\vec{x},0) =0\)) so that the initial ion gyro-center densities fulfill the condition
\begin{align}
\label{eq:init1}
 \ne(\vec{x},0)  &= \Gamma_{1,i} \Ni(\vec{x},0) .
\end{align}
\\
The  initial ion entropy and  potential energy is  \(S_i(0)\approx S_e(0)\) and \(H_i(0) \approx0\), respectively. 
Further, the initial ion kinetic energy is vanishing \(E_{k,i}(0) = 0\).
\subsubsection{Rotating Gaussian}
For the second initial condition the total polarization density is constant \(\vec{P}_1(\vec{x},0)  + \vec{P}_2(\vec{x},0)  = 
\vec{const.}\), which gives rise to an initial electric field for finite ion temperature.
The constant polarization density condition is fullfilled by 
\begin{align}
\label{eq:init2}
 \ne(\vec{x},0)  &=  \Ni(\vec{x},0). 
\end{align}
In the long wavelength limit we obtain the initial \ExB rotation is \(\vec{\nabla}_\perp \phi \approx \Tip/(2 q_i) \vec{\nabla}_\perp \ln n + \vec{const.}\). Thus the sum of the \ExB and diamagnetic vorticity density is non-vanishing and no force balance is established initially~\cite{held2016,wiesenberger2020}.
\\
From Eq.~\eqref{eq:init2} follows that the initial ion entropy and potential energy is given by  \(S_i(0)= S_e(0)\) and  \(H_i(0) = H_e(0)\), respectively. 
Further, the initial ion kinetic energy is finite \(E_{k,i}(0) 
\approx  \frac{m_i c_{s0}^4}{4 \neref \Omega_0^2}\left(\frac{ \Tip}{\tep}\right)^2 
 \pi g(\Delta \ne) \), where we used the long wavelength limit expression, 
\(\vec{\nabla}_\perp \phi = \Tip/(2 q_i) \vec{\nabla}_\perp \ln \Ni\), assumed small variations in the magnetic field magnitude 
\(B^{-2}\approx B_0^{-2}\) and defined \(g(x):=(6 x - \pi^2 + 3 
\ln(1+x)\ln((1+x)/x^2) + 6 \text{Li}_2 (1/(1+x)))/3\). 
For small fluctuation amplitudes the initial ion kinetic energy  quadratically depends on the relative fluctuation amplitude 
\(E_{k,i}(0)  \approx  \frac{m_i c_{s0}^4}{8 \neref 
\Omega_0^2}\left(\frac{ \Tip}{\tep}\right)^2 \pi (\Delta \ne)^2 \). 
Note that the initial entropies are much larger than the initial ion kinetic energy, \(\sum_s \Tzp S(0)\ll - E_{k,i}(0)\),  as long as 
\(\sigma^2/\rho_{s0}^2 \gg (\Tip/\tep)^2 /\left[4(1+\Tip/\tep)\right]\). This condition is fulfilled for all parameters that we consider. Consequently, we neglect the contribution of the initial ion kinetic energy in our further analysis.

\section{Unified scaling laws for interchange blob dynamics}
\label{sec:scalinglaws}
\begin{figure}[ht]
\centering
\includegraphics[width=0.475\textwidth]{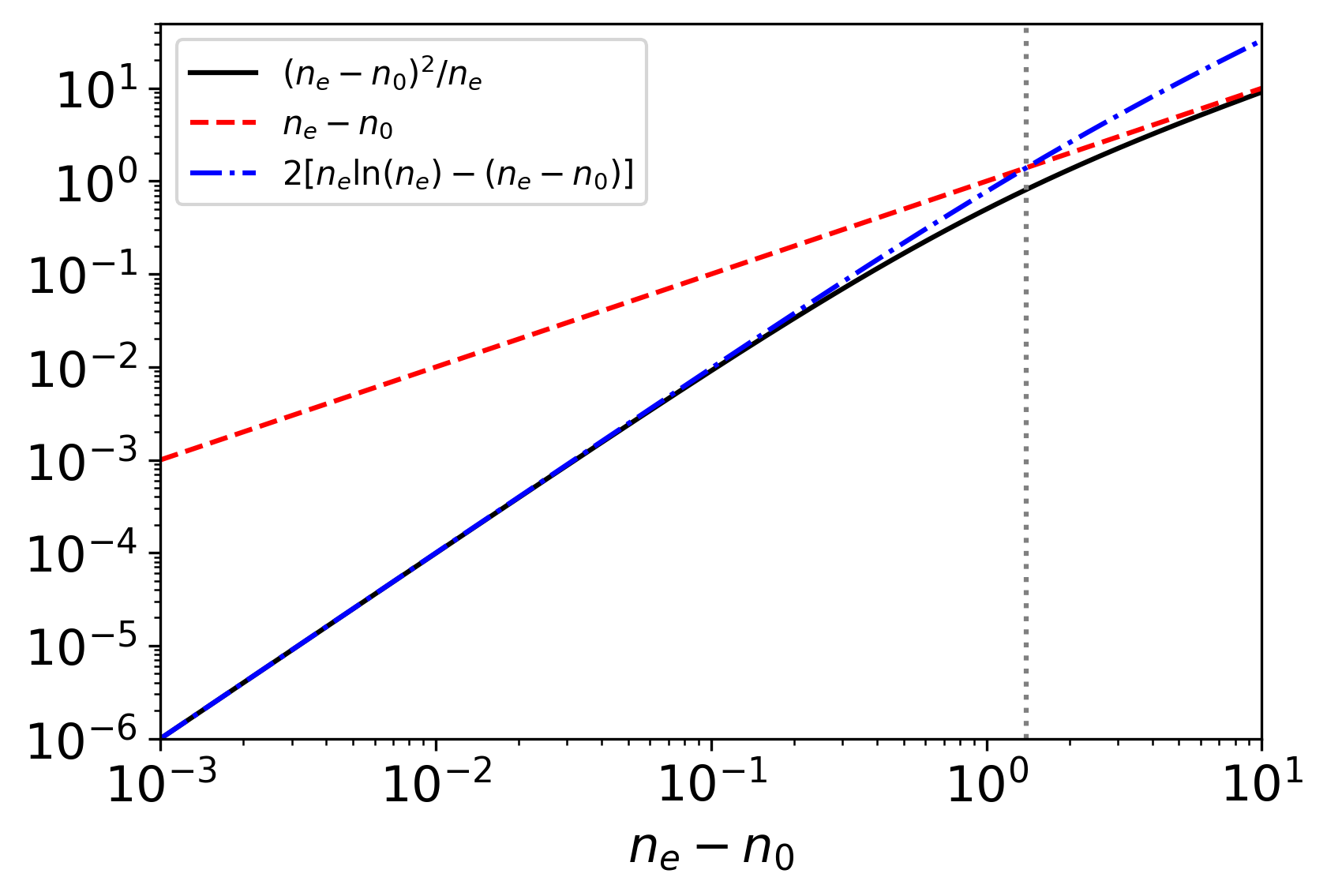}
\caption{The upper bounds that result into the two integral estimates of Eq.~\eqref{eq:integralestimate} are depicted. The dotted line represents the critical density \(n_{e,c}\approx2.397\neref\) of the two estimates.}
\label{fig:Scaling_estimate}
\end{figure}

We define three measures for the interchange blob dynamics. These are the center of mass position, velocity and acceleration
\begin{subequations}
\begin{align}
    \vec{X} &:= M_e^{-1} \int dA \vec{x} (\ne-\neref), \\
    \vec{V} &:= \frac{d \vec{X}}{d t} ,\\
    \vec{A} &:= \frac{d \vec{V}}{d t}.
\end{align}
\end{subequations}
The x-component of the center of mass position is related to the potential energy of the electrons via \(H_e(t) \approx -M_e T_e (X-x_0)/R_0\). Thus, the potential energy is decreasing in time \(\sum_s H(t)\leq \sum_s  H(0)\) for blobs (and holes). 
This allows us to show that the kinetic energy as well as the entropy is increasing in time \(\sum_s E_k(t) \geq\sum_s E_k(0)\) and \(\sum_s \Tzp S(t)  \geq\sum_s \Tzp S(0)\), respectively. Thus, the second law of thermodynamics is fulfilled.
\\
In the following we will deduce scaling laws for the x-component of the center of mass acceleration \(A_x\) and velocity \(V_x\) as well as for the interchange growth rate \(\gamma\). We base these scaling laws on the following estimate for the squared center of mass momentum~\cite{wiesenberger2017}
\begin{align}
 (M_e V_{x})^2 
 &= \left(\int d A \ne B^{-1}\partial_y \phi \right)^2 \nonumber \\
 &= \left(\int d A (\ne-\neref) B^{-1}\partial_y \phi\right)^2 \nonumber \\
  &\leq \int d A (\ne-\neref)^2/\ne \int d A \ne (B^{-1}\partial_y \phi)^2 
\nonumber \\
  &\leq \frac{2}{m_i} D(t) E_{k,i}(t) ,
  \label{eq:Vx_estimate}
\end{align}
where we use the Cauchy-Schwartz inequality and the estimate \(\int d A \ne (B^{-1}\partial_y \phi)^2 \leq 2 E_{k,i}(t)/m_i\). 
Further, we introduce  the integral \(D(t) := \int d A (\ne-\neref)^2/\ne\). For blobs (\(\ne \geq n_{e0}\)) there exist two close upper bounds for the expression inside the latter integral, which 
are depicted in Fig.~\ref{fig:Scaling_estimate}. These are the (scaled) electron entropy density \(2\left[\ne \ln (\ne/\neref) - (\ne-\neref)  \right]\) for \(\neref\leq \ne< n_{e,c} \) and the relative density \(\ne-\neref\) for \( \ne\geq n_{e,c}\). Here, we defined the critical density \(n_{e,c}:=\left\{1+\exp{\left[ 3/2+W\left(-3/2/\exp{(3/2)}\right) \right] }\right\}\neref\approx2.397\neref\) 
with the product logarithm  \(W(x)\), which follows from equating the scaled electron entropy and relative density.
Thus, we can further assess the integral \(D(t)\) to
\begin{align}
\label{eq:integralestimate}
 D(t)\leq
 \begin{cases} 
 -2 S_e(0),\ \mathrm{if} \ \neref\leq \ne< n_{e,c}  \\
M_e(0), \ \ \ \ \mathrm{if} \ \ne\geq n_{e,c}   
\end{cases},
\end{align}
where we used the identities \(S_e(t)\geq S_e(0)\) and \(M_e(t)=M_e(0)\).

\subsection{Acceleration}
From the conservation of Eq.~\eqref{eq:Z} results
\(E_{k,i}(t) = \sum_s \left[H(0)- H(t)\right] +E_{k,i}(0)\approx (\tep +\Tip) M_e 
X/R_0\). Together with the first estimate of Eq~\eqref{eq:integralestimate} in 
Eq.~\eqref{eq:Vx_estimate} we obtain
\begin{align}
\label{eq:Vx_acc_estimate}
 (M_e V_{x})^2 &\leq -\frac{c_{s}^2 4 S_e(0)  M_e(t) X(t)}{R_0},
\end{align}
with the ion acoustic speed \(c_s:=\sqrt{(\tep+\Tip)/m_i}\) and its cold ion limit equivalent \(c_{s0} := \sqrt{\tep/m_i}\).
\\
Assuming now a linear acceleration \(V_x = A_x t  \) and \(X(t) = A_x t^2/2\) in 
Eq.~\eqref{eq:Vx_acc_estimate} results into an estimate for the acceleration~\cite{kube2016,wiesenberger2017}
\begin{align}
\label{eq:acceleration}
 A_x &= -\mathcal{Q} \frac{c_{s}^2}{R_0}\frac{2 S_e(0)}{M_e}
\nonumber \\
     &\approx \frac{\mathcal{Q}}{2} \frac{c_{s}^2}{R_0} \frac{ \Delta \ne}{\neref + 2/9 
\Delta \ne}.
\end{align}
Here, we introduced the parameter \(\mathcal{Q}\in \left(0,1\right]\) and used 
the (1,1) Pad{\'e} approximation for \(\frac{S_e(0)}{M_e}\approx -\frac{1}{4}\frac{\Delta \ne}{\neref + 2/9 
\Delta \ne}\).

Analogously, we obtain with the second estimate of Eq~\eqref{eq:integralestimate} the upper bound for the 
acceleration \( A_x\leq c_{s}^2/R_0\). This upper bound can be interpreted as an effective gravity \(g_\mathrm{eff}:= c_{s}^2/R_0\) that would arise when replacing the \(\vec{\nabla} B\) drift with a gravitational drift.
\subsection{Velocity}
\subsubsection{Linear velocity scaling}
From the conservation of energy (Eq.~\eqref{eq:F}) results that the kinetic energy of the ions is bounded by \( E_{k,i}(t)\leq  E_{k,i}(t) -\sum_{s} \Tzp S(t) 
= E_{k,i}(0) - \sum_s \Tzp S(0)\approx -(\tep + \Tip) S_e(0)\). This together 
with the first estimate of Eq~\eqref{eq:integralestimate} in 
Eq.~\eqref{eq:Vx_estimate} yields the linear velocity scaling~\cite{kube2016,wiesenberger2017}
\begin{align}
\text{max}(V_x)  &=- \mathcal{Q} c_{s} \frac{2 
S_e(0)}{M_e(0)} 
 \nonumber \\
 &\approx \frac{\mathcal{Q}}{2} c_{s}   \frac{\Delta \ne}{\neref + 2/9 
\Delta \ne} \nonumber \\
\label{eq:linearscaling}
 &\approx \frac{\mathcal{Q}}{2}  c_{s}   \frac{\Delta \ne}{\neref},
\end{align}
where the Pad{\'e} approximation is Taylor expanded for small relative density fluctuations.
\subsubsection{Square root velocity scaling}
The square root scaling can be deduced 
by taking the time derivative of the 
\fullF Poisson Eq.~\eqref{eq:PoissonFF} and applying the long wavelength approximation~\cite{held2016,gerru2022}
\begin{align}
\label{eq:dtvor}
 \frac{\partial}{\partial t} \mathcal{W} + \vec{\nabla}\cdot \vec{\nabla}\cdot \left(\vec{\omega} \vec{u}_E\right) &= \frac{c_s^2}{R_0} \partial_y \ne.
\end{align}
where we introduced the vorticity density \(\mathcal{W} := \bhat \cdot \nablav \times\vec{j} = \vec{\nabla}\cdot \vec{\omega} \) and the vector
\( \vec{\omega} := -\bhat \times \vec{j}\) with the sum of the \ExB and ion diamagnetic current density \(\vec{j} :=\ne \left(\vec{u}_E+\vec{u}_{D,i}\right) \) and the ion diamagnetic drift \(\vec{u}_{D,i}:= \frac{\bhat \times \vec{\nabla} (\ne \Tip)}{q_i \ne B}\). 
A scale analysis of Eq.~\eqref{eq:dtvor} yields the square root velocity scaling~\cite{ott1978,garcia2005,madsen2011,held2016}
\begin{align}
\label{eq:squarerootscalingBoussinesq}
 \text{max}(V_x)&= 
\mathcal{R} c_s \sqrt{\frac{\sigma}{R_0}\frac{\Delta \ne}{\neref  }}   ,
\end{align}
with parameter \(\mathcal{R}\in\left(0,1\right]\).
\\
Interestingly, a square root scaling results also from similar methods as we used for the linear velocity scaling law (Eq.~\eqref{eq:linearscaling}). To this end, we use the same bound \( E_{k,i}(t)\leq  E_{k,i}(t) - \sum_{s}\Tzp S(t) 
=
 E_{k,i}(0) - \sum_s \Tzp S(0)\approx -(\tep + \Tip) S_e(0)\) together with the 
second estimate of Eq~\eqref{eq:integralestimate} in Eq.~\eqref{eq:Vx_estimate}.
This yields 
\( \text{max}(V_x) = 
\mathcal{R}c_s\sqrt{\frac{\sigma }{R_0}\frac{\Delta \ne}{\neref + 2/9 
\Delta \ne}} \)  where \(\mathcal{R}\in\left(0,R_0/(\sigma 
\sqrt{2})\right]\) is estimated through the scale analysis expression of Eq.~\eqref{eq:squarerootscalingBoussinesq}.
This scaling law agrees in the Oberbeck-Boussinesq limit with  Eq.~\eqref{eq:squarerootscalingBoussinesq}. However, in the cold ion limit the numerical data clearly supports the square root velocity scaling law of Eq.~\eqref{eq:squarerootscalingBoussinesq} and does not justify the inclusion of the  non-Oberbeck-Boussinesq factor.
\subsubsection{Unified velocity scaling}
Following similar methods as in Ref.~\cite{pecseli2016, wiesenberger2017} we can combine the linear and square root velocity scaling law into a unified scaling law that accounts for finite ion temperatures
\begin{align}
\label{eq:Vx_unified}
 \text{max}(V_x)  &= 
\frac{\mathcal{R}^2}{\mathcal{Q}}c_s\frac{\sigma}{R_0} 
\left[\sqrt{1+\left(\frac{\mathcal{Q}}{\mathcal{R}}\right)^2 \frac{\Delta 
\ne }{ \neref} \frac{R_0}{\sigma}} -1\right].
\end{align}
The latter resembles the small and large relative density fluctuation limits, Eq.~\eqref{eq:linearscaling} respectively Eq.~\eqref{eq:squarerootscalingBoussinesq}, correctly. 
Note that the only difference to the proposed unified scaling law in Ref.~\cite{wiesenberger2017} is the factor \(\sqrt{1+\Tip/\tep}\) that is inherent in the ion acoustic speed \(c_s\).
\subsection{Interchange growth rate}
The time at which the maximum velocity occurs is estimated by
the inverse  interchange growth rate \(t_{\text{max}(V_x)} \sim \gamma^{-1}\), 
which we define as
\begin{align}
\label{eq:growthrate}
 \gamma := A_x/\text{max}(V_x).
\end{align}
\section{Numerical experiments} \label{sec:numerical}
The full-F gyro-fluid model and its three approximations (long wavelength, Oberbeck-Boussinesq, long wavelength + Oberbeck-Boussinesq) is numerically solved by using the open source library \textsc{Feltor}~\cite{wiesenberger2019,feltor}. We choose
a discontinuous Galerkin discretization on a rectangular grid in space and use an explicit adaptive timestepper based on the Bogacki-Shampine embedded Runge-Kutta method in time. 
The convective terms
in Eq.~\eqref{eq:continuityFF} are thus discretized using
a discontinuous Galerkin upwind scheme~\cite{cockburn-shu-2001}, while the elliptic polarization equation~\eqref{eq:PoissonFF} is discretized using local discontinuous Galerkin methods~\cite{cockburn2001}. 
The advantages of these methods are their high order, low numerical diffusion and ease of parallelization. We thus take advantage of Nvidia's V100 GPUs on the Marconi M100 supercomputer for all our simulations.

We remark that the numerical application of the \(\sqrt{\Gamma_0}\) operation demands to efficiently solve a matrix function equation of the form $\sqrt{A}x = b$, where $A$ is the discrete form of $1-\rho^2 \Delta_\perp$ in configuration space and the matrix function is the matrix square root.
Our approach is based on a Krylov approximation in order to drastically reduce the original matrix size for the matrix function (i.e. square root function) computation by projecting onto a Krylov subspace.
More specifically, we use the symmetric Lanczos algorithm
to obtain a matrix decomposition of $A$ in tridiagonal form. The matrix function is then applied to the small tridiagonal matrix in terms of either an eigenvalue decomposition or a Cauchy-contour integral~\cite{hale08}. This is by far more efficient than for example an eigenvalue decomposition of the original large sparse matrix $A$. The remaining problem is the construction of a judicious stopping criterion for the Lanczos algorithm. For this  the recently proposed solution  of Ref.~\cite{eshghi2021} is utilized.
We remark that all of our implementations are freely available~\cite{feltor} and well-documented. We employ unit and
integration tests and we observe the expected order of convergence in all manufactured test problems. An in depth discussion of the implementation and application of matrix-function computation will be given in a separate manuscript.
The supplemental dataset to this contribution ensures bitwise reproducibility of the herein reported results~\cite{dataset}.

We resolve all herein presented simulations with \(n=5\) polynomial coefficients and \(N_x=N_y=300\) grid cells with a box size of \(L_x=L_y=40 \sigma\). In total this amounts to approximately $2$ million grid points. This high order and  fine resolution is necessary to resolve very small scale \ExB vorticity structures in the numerical solutions as will become evident in this section. Boundary conditions are Dirichlet in x-direction and periodic in y-direction. For the sake of regularizing the resulting $5$-th order upwind discretization we add a very small hyperdiffusion term of 2nd order to the continuity equations. The mass diffusion coefficient \(\nu\) is determined through the Rayleigh number \(\mathrm{Ra} := \frac{g_\mathrm{eff} \sigma^3 \Delta \ne}{\nu^2 \neref}\), which is fixed to \(\mathrm{Ra} = 2\times 10^9\). The Schmidt number is set to unity so that the viscosity coefficient equals the mass diffusion coefficient. As a consequence, the chosen parameters span the turbulent regime~\cite{garcia2005,kube2012}.
\\
The chosen physical parameter space lies in a typical experimental regime, but also in a regime  where the approximations in question are expected to fail. This is for small blobs with blob size \(\sigma=5\rho_{s0}\), hot ion temperature \(\Tipref/\tepref =4\) and density perturbations that  encompass values in the range \(\Delta \ne /n_{e0}\in \left[10^{-3}, 10\right]\). The parameter for the magnetic field gradient is chosen to \(\frac{\rho_{s0}}{R_0} = 0.00015\) to allow direct comparison to previous studies~\cite{held2016}.
\subsection{Fundamental dynamics of cold and hot blobs}~\label{sec::coldhot}
\begin{figure*}
\centering
\includegraphics[width= 1.0\textwidth]{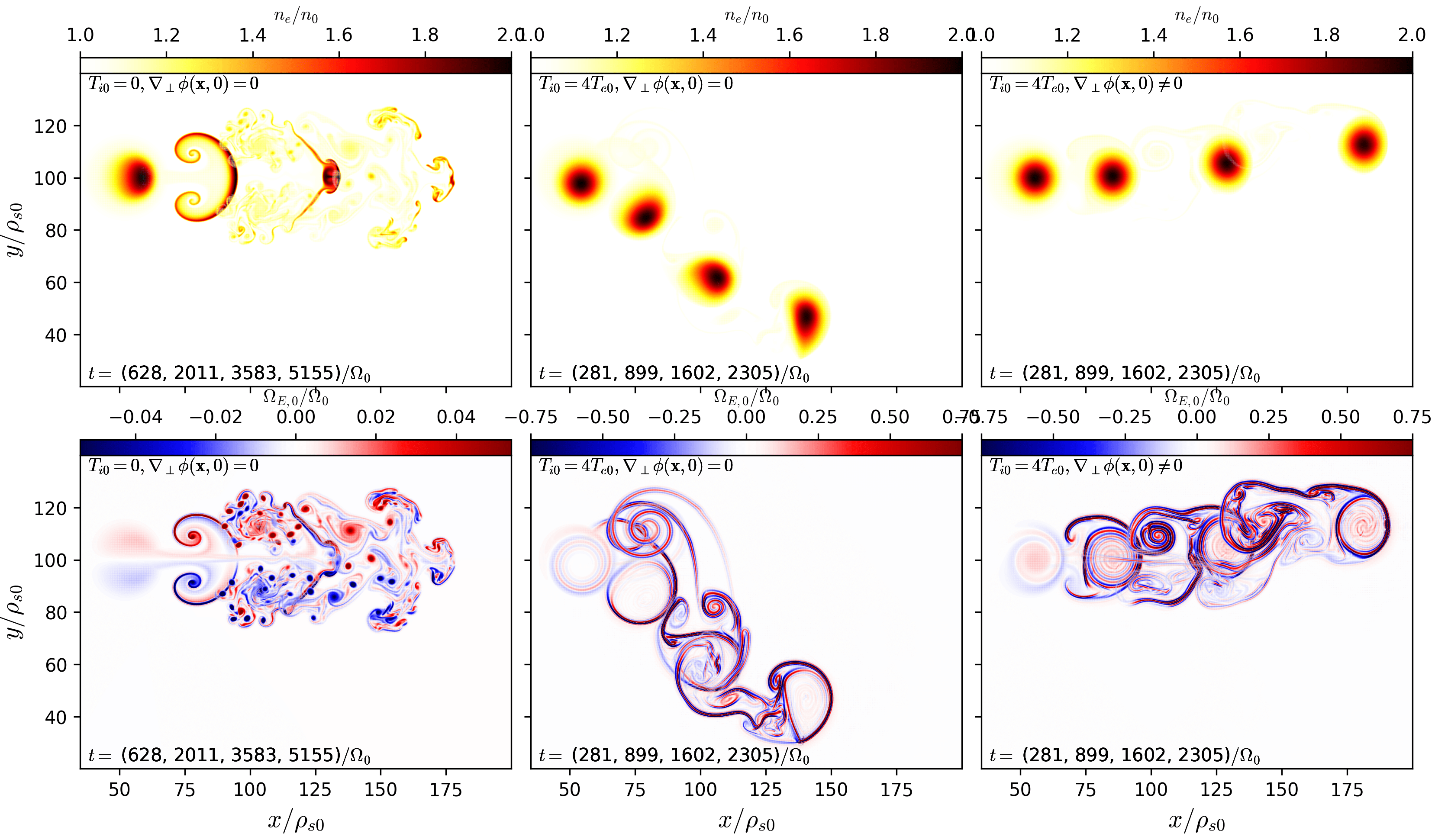}
\caption{The evolution of the electron density \(\ne\) and the \ExB vorticity  is shown for two different ion to electron temperature ratios \(\Tip/\tep=\left(0.0,4.0\right) \) and two different initial conditions for the electric field. The initial blob amplitude and blob size are \(\Delta \ne/\neref=1.0\) and \(\sigma=5\rho_{s0}\), respectively. The respective movies (\href{run:./tau0-ff-A1.mp4}{"Full-F, amp=1, tau=0"}, 
\href{run:./tau4-ff-A1.mp4}{"Full-F, amp=1, tau=4"}, 
\href{run:./tau4-ff-A1-0pol.mp4}{"Full-F, amp=1, tau=4, 0pol"})  to this figure can be found in the supplementary material.}
 \label{fig:coldhotblobs}
\end{figure*}

It is well known that the finite ion temperature effects lead to fundamentally different blob dynamics in comparison to cold ion temperature~\cite{madsen2011,wiesenberger2014,held2016}, which are visualized in Fig.~\ref{fig:coldhotblobs}. The respective movies (\href{run:./tau0-ff-A1.mp4}{"Full-F, amp=1, tau=0"}, 
\href{run:./tau4-ff-A1.mp4}{"Full-F, amp=1, tau=4"}, 
\href{run:./tau4-ff-A1-0pol.mp4}{"Full-F, amp=1, tau=4, 0pol"}) to this figure can be found in the supplementary material.
\\
In the cold ion limit a nearly up-down symmetric mushroom like blob develops that is advected purely into the x-direction~\cite{garcia2005}. The mushroom like structure consists of a steepening blob front and lobes that roll itself up. This structure quickly disintegrates after the initial linear acceleration phase.
The x-directed \ExB propagation is a consequence of a dipole in the electric field that emerges due to the interplay of polarization and the \(\vec{\nabla} B\) drift. 
\\
By contrast, for finite ion temperature the blob strongly retains its initial shape both for the non-rotating and rotating Gaussian initial condition. The tendency to maintain its initial shape is attributed to small scale \ExB shear flows~\footnote{Resolving these small scale \ExB vorticity structures in numerical simulations is challenging, requiring numerical schemes with low artificial diffusion or vast numerical resolution.} that suppress the removal of mass via small eddy-satellites. These shear flows can be attributed to the gyro-amplification discussed in Section~\ref{sec:amplification}. The initial dipole in the vorticity rolls
up into a spiral of strongly sheared flows. 
These are strongest at the edge of the blob and decrease in magnitude towards the blob center. The small scale blob \ExB shear flows share features of zonal flows of magnetized plasma turbulence~\cite{held2018}. The emerging blob shape is most reminiscent of a jellyfish, in particular for the rotating Gaussian initial condition~\cite{held:phdthesis:17}.
The additional blob motion in y-direction strongly depends on the initial condition. In particular the blob is propagating also into the negative y-direction for the non-rotating initial condition, while the y-motion mostly disappears for an initially rotating Gaussian.
\subsection{Arbitrary wavelength polarization and non-Oberbeck-Boussinesq effects  on nonlinear blob dynamics}~\label{sec:fullkNOB}
\begin{figure*}
\centering
\includegraphics[width= 1.0\textwidth]{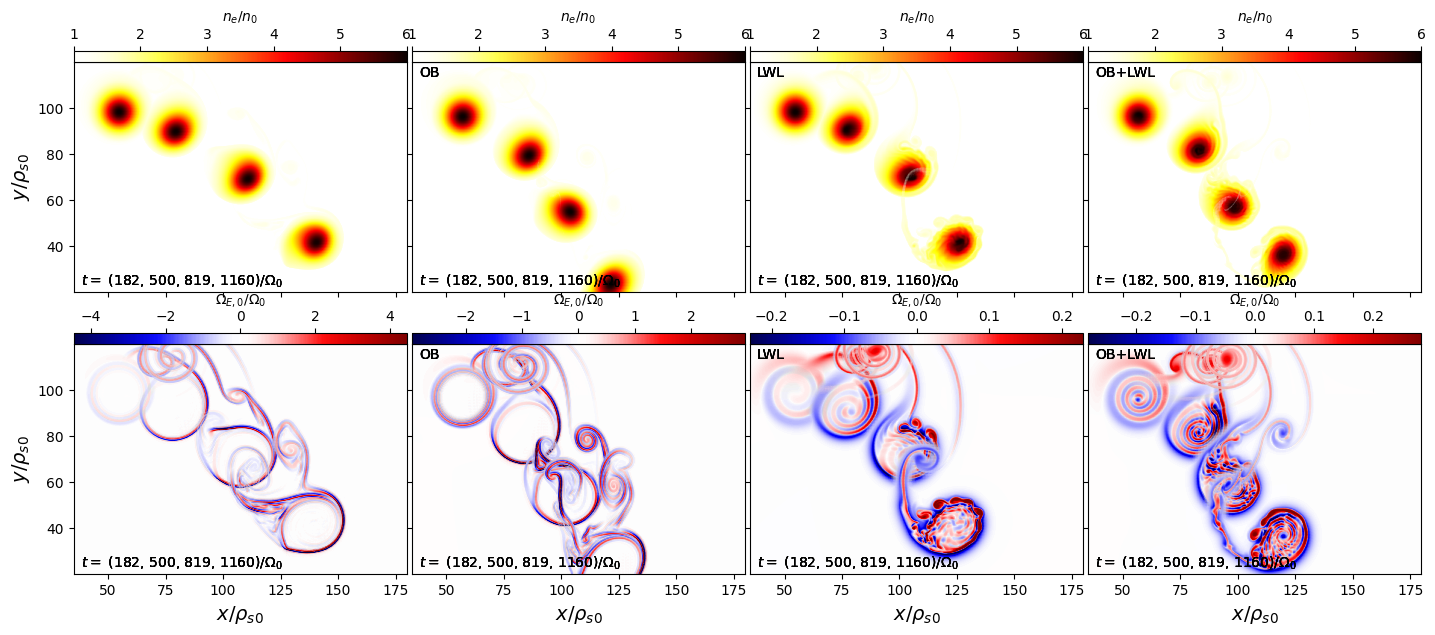}
\caption{The evolution of the electron density \(\ne\) and the \ExB vorticity \(\Omega_E\)  is shown for the very large initial blob amplitude \(\Delta \ne/\neref=5.0\), blob size \(\sigma=5 \rho_{s0}\) and ion to electron temperature ratio \(\Tip/\tep=4\). Various model simplifications are shown from left to right (cf. Table~\ref{table:poltreatment}). Note the change in vorticity pattern and magnitude. The respective movies
(\href{run:./tau4-ff-A5.mp4}{"Full-F, amp=5, tau=4"}, 
\href{run:./tau4-df-A5.mp4}{"OB, amp=5, tau=4"}, 
\href{run:./tau4-ff-lwl-A5.mp4}{"LWL, amp=5, tau=4"}, 
\href{run:./tau4-df-lwl-A5.mp4}{"OB+LWL, amp=5, tau=4"}) to this figure can be found in the supplementary material.}
\label{fig:blobsA5}
\end{figure*}

\begin{figure*}
\centering
 \includegraphics[width= 1.0\textwidth]{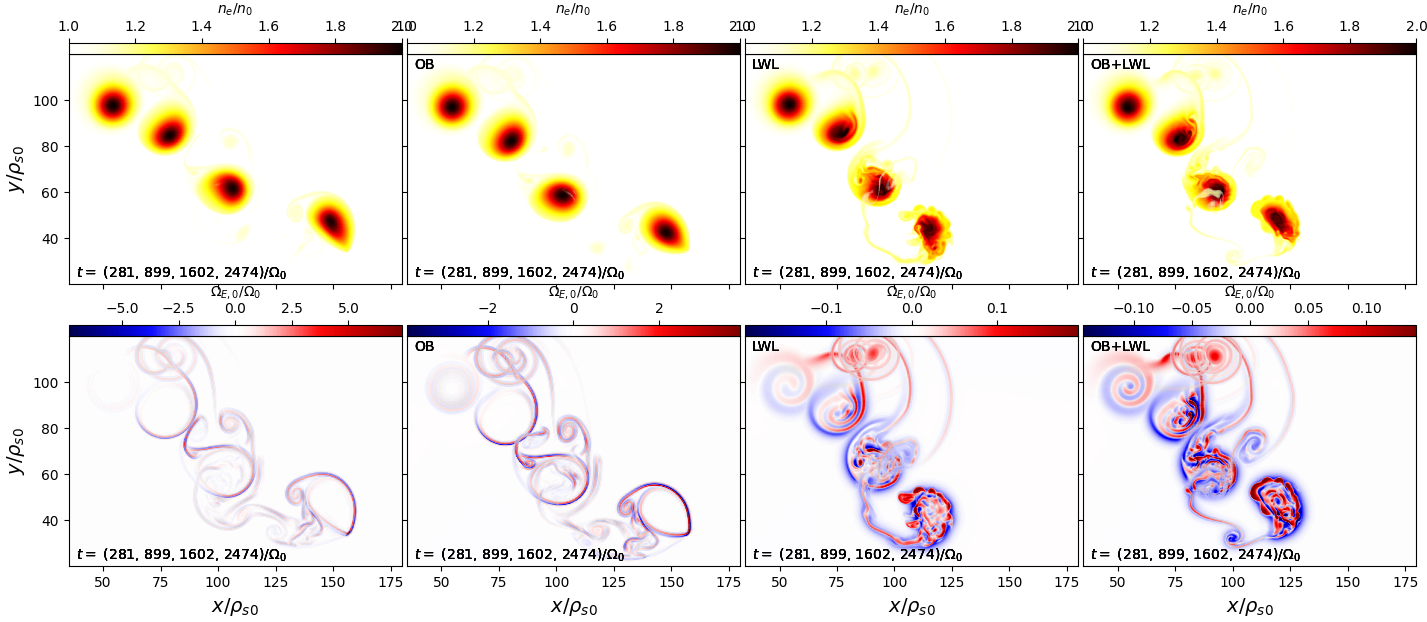}
\caption{The evolution of the electron density \(\ne\) and the \ExB vorticity \(\Omega_E\)  is shown for large initial blob amplitude \(\Delta \ne/\neref=1.0\), blob size \(\sigma=5 \rho_{s0}\) and ion to electron temperature ratio \(\Tip/\tep=4\). Various model simplifications are shown from left to right (cf. Table~\ref{table:poltreatment}). Note the change in vorticity pattern and magnitude. 
The respective movies
(\href{run:./tau4-ff-A1.mp4}{"Full-F, amp=1, tau=4"}, 
\href{run:./tau4-df-A1.mp4}{"OB, amp=1, tau=4"}, 
\href{run:./tau4-ff-lwl-A1.mp4}{"LWL, amp=1, tau=4"}, 
\href{run:./tau4-df-lwl-A1.mp4}{"OB+LWL, amp=1, tau=4"}) to this figure can be found in the supplementary material. }
 \label{fig:blobsA1}
\end{figure*}

\begin{figure*}
\centering
 \includegraphics[width= 1.0\textwidth]{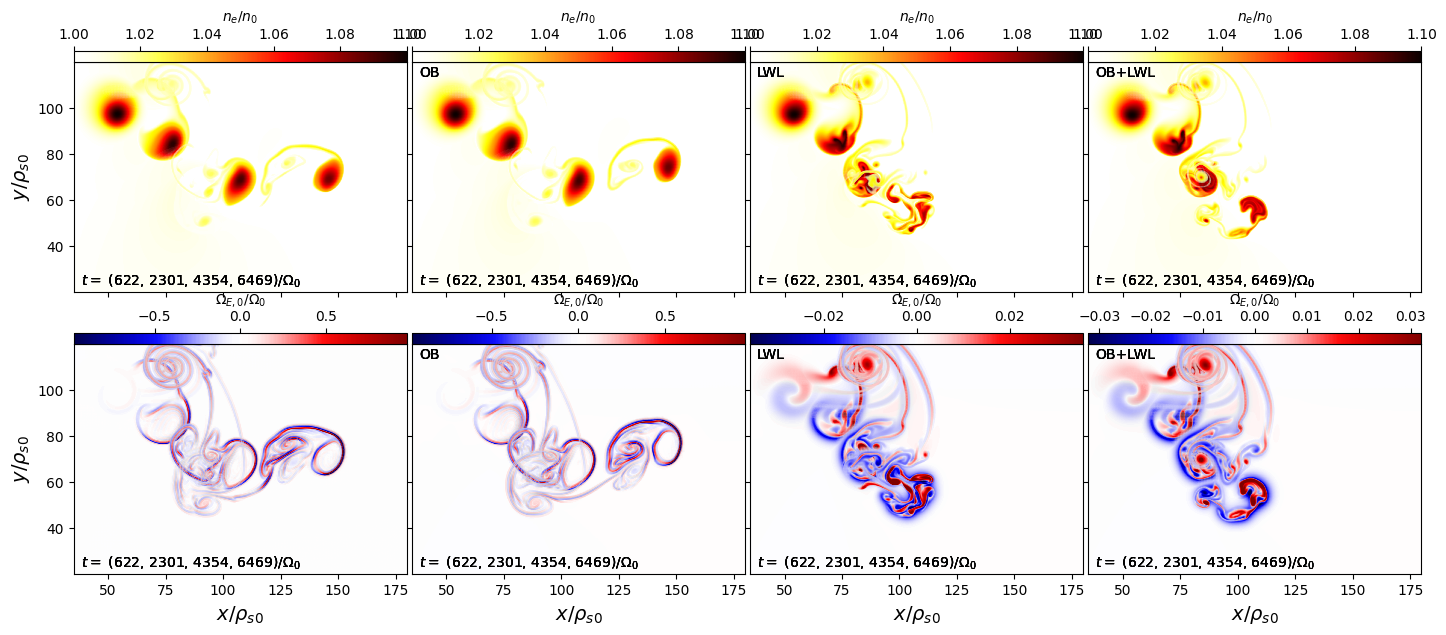}
 \caption{The evolution of the electron density \(\ne\) and the \ExB vorticity \(\Omega_E\)  is shown for the small initial blob amplitudes \(\Delta \ne/\neref=0.1\), blob size \(\sigma=5 \rho_{s0}\) and ion to electron temperature ratio \(\Tip/\tep=4\). Various model simplifications are shown from left to right (cf. Table~\ref{table:poltreatment}). Note the change in vorticity pattern and magnitude. The respective movies 
 (\href{run:./tau4-ff-A01.mp4}{"Full-F, amp=0.1, tau=4"}, 
\href{run:./tau4-df-A01.mp4}{"OB, amp=0.1, tau=4"}, 
\href{run:./tau4-ff-lwl-A01.mp4}{"LWL, amp=0.1, tau=4"}, 
\href{run:./tau4-df-lwl-A01.mp4}{"OB+LWL, amp=0.1, tau=4"}) to this figure can be found in the supplementary material.}
 \label{fig:blobsA01}
\end{figure*}

We now turn our attention to the differences between the considered models (cf. Table~\ref{table:poltreatment}) for various initial blob amplitudes and for the non-rotating Gaussian initial condition (Eqs~\eqref{eq:init_ne} and~\eqref{eq:init1}). In Figures~\ref{fig:blobsA5},~\ref{fig:blobsA1}~\ref{fig:blobsA01} the temporal blob evolution
is shown for three typical blob amplitudes \(\Delta \ne/\neref=(5.0,1.0,0.1)\). The respective movies
(\href{run:./tau4-ff-A5.mp4}{"Full-F, amp=5, tau=4"}, 
\href{run:./tau4-df-A5.mp4}{"OB, amp=5, tau=4"}, 
\href{run:./tau4-ff-lwl-A5.mp4}{"LWL, amp=5, tau=4"}, 
\href{run:./tau4-df-lwl-A5.mp4}{"OB+LWL, amp=5, tau=4"}, 
\href{run:./tau4-ff-A1.mp4}{"Full-F, amp=1, tau=4"}, 
\href{run:./tau4-df-A1.mp4}{"OB, amp=1, tau=4"}, 
\href{run:./tau4-ff-lwl-A1.mp4}{"LWL, amp=1, tau=4"}, 
\href{run:./tau4-df-lwl-A1.mp4}{"OB+LWL, amp=1, tau=4"}, 
\href{run:./tau4-ff-A01.mp4}{"Full-F, amp=0.1, tau=4"}, 
\href{run:./tau4-df-A01.mp4}{"OB, amp=0.1, tau=4"}, 
\href{run:./tau4-ff-lwl-A01.mp4}{"LWL, amp=0.1, tau=4"}, 
\href{run:./tau4-df-lwl-A01.mp4}{"OB+LWL, amp=0.1, tau=4"}) that highlight the differences between the considered models can be found in the supplementary material.
\\
For all models we observe the initial rolling up or spiraling in the \ExB vorticity \(\Omega_E:= \bhat \cdot \nablav \times \vec{u}_E\) and the consequent x- and y-directed blob motion~\cite{madsen2011,wiesenberger2014,held2016}. The amplitude of the \ExB shear flows decreases by roughly an order of magnitude  if the long wavelength approximation is applied. Further, for the long wavelength approximated models the \ExB shear flows are at much larger scales and no longer show up a clear gradient in magnitude from the blob center towards the blob edge. As a consequence, the long wavelength approximation leads to less coherent blobs that disintegrate more quickly. The reduction in blob coherence is most pronounced for small amplitudes and weakens with increasing initial blob amplitude.
\\
Apart from the effect of blob shape and structure the blob propagation is also significantly affected by the considered approximations. 
The Oberbeck-Boussinesq approximation leads to an increased blob displacement for very large blob amplitudes (\(\Delta \ne/\neref=(5.0,1.0)\) in Fig~\ref{fig:blobsA5} and ~\ref{fig:blobsA1}). This is best visible at the end of the linear acceleration phase (roughly coinciding with the second time point in Fig~\ref{fig:blobsA5} and ~\ref{fig:blobsA1}).
After the linear acceleration phase the long wavelength approximation results in different blob positions due to the dissociation of smaller eddy-satellites that can change the movement of the blob or due to complete disintegration of the blob.
This is most clearly recognizable for the small amplitude blob (\(\Delta \ne/\neref=0.1\) in Fig~\ref{fig:blobsA01}).
\\
Finally, we remark that the rotating Gaussian initial condition is not included in this analysis, since in this case one of the physical initial fields, specifically the electric potential \(\phi(\vec{x},0)\), changes if the long wavelength or Oberbeck-Boussinesq approximation is applied. However, we observed that the long wavelength approximation results in less coherent blobs in comparison to the non-rotating Gaussian initial condition. This underpins the results that are obtained for the non-rotating Gaussian intial condition.
\subsection{Blob compactness}~\label{sec:blobcompactness}
\begin{figure}
\centering
\includegraphics[width=0.475\textwidth]{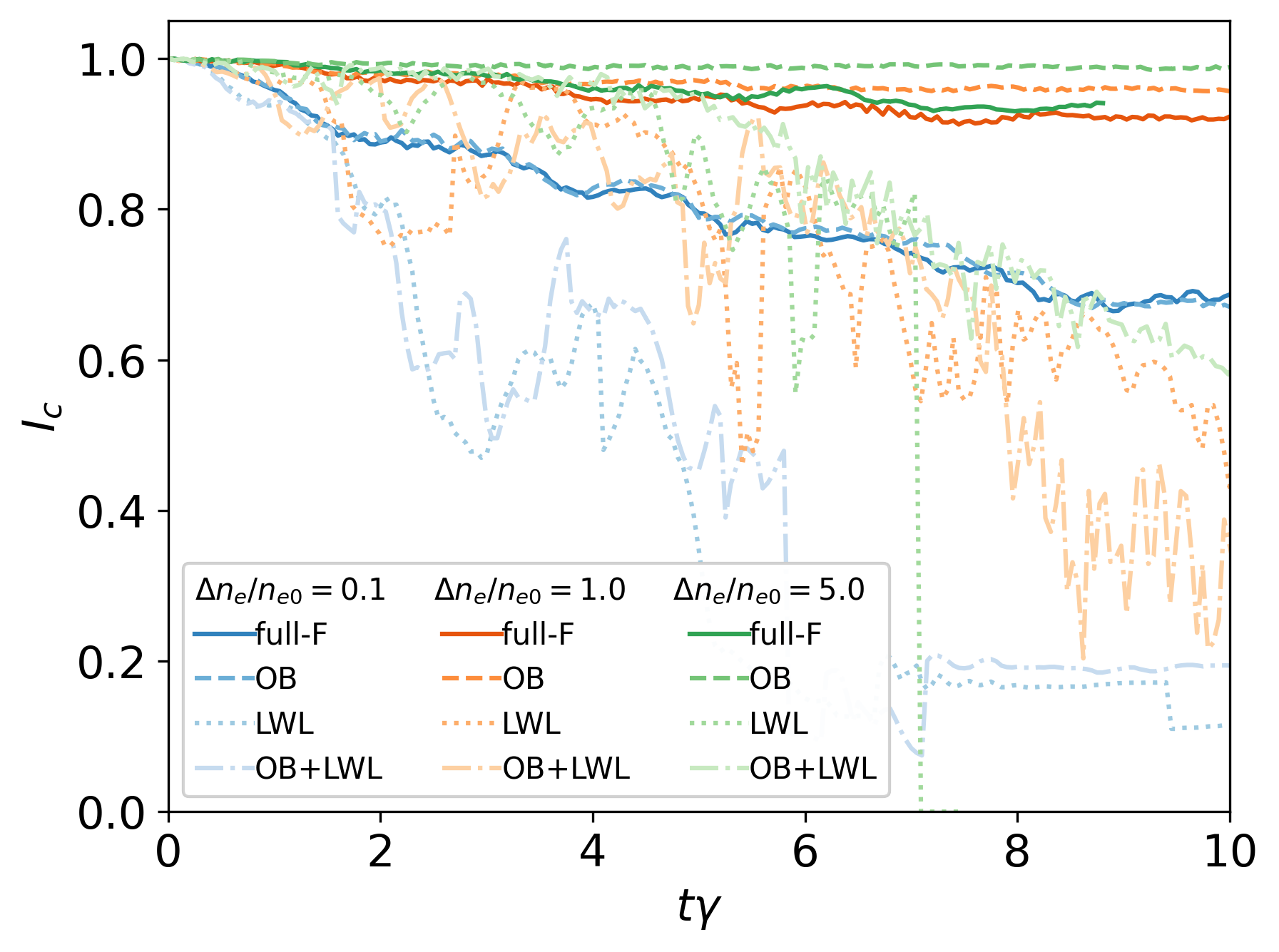}
\caption{The time evolution of the blob compactness is shown for three characteristic blob amplitudes \(\Delta \ne/\neref = \left(5.0, 1.0,0.1 \right)  \), all considered models (Table~\ref{table:poltreatment}) and the non-rotating Gaussian initial conditions.}
\label{fig:IcVStgamma}
\end{figure}

\begin{figure}[ht]
\centering
\includegraphics[width=0.475\textwidth]{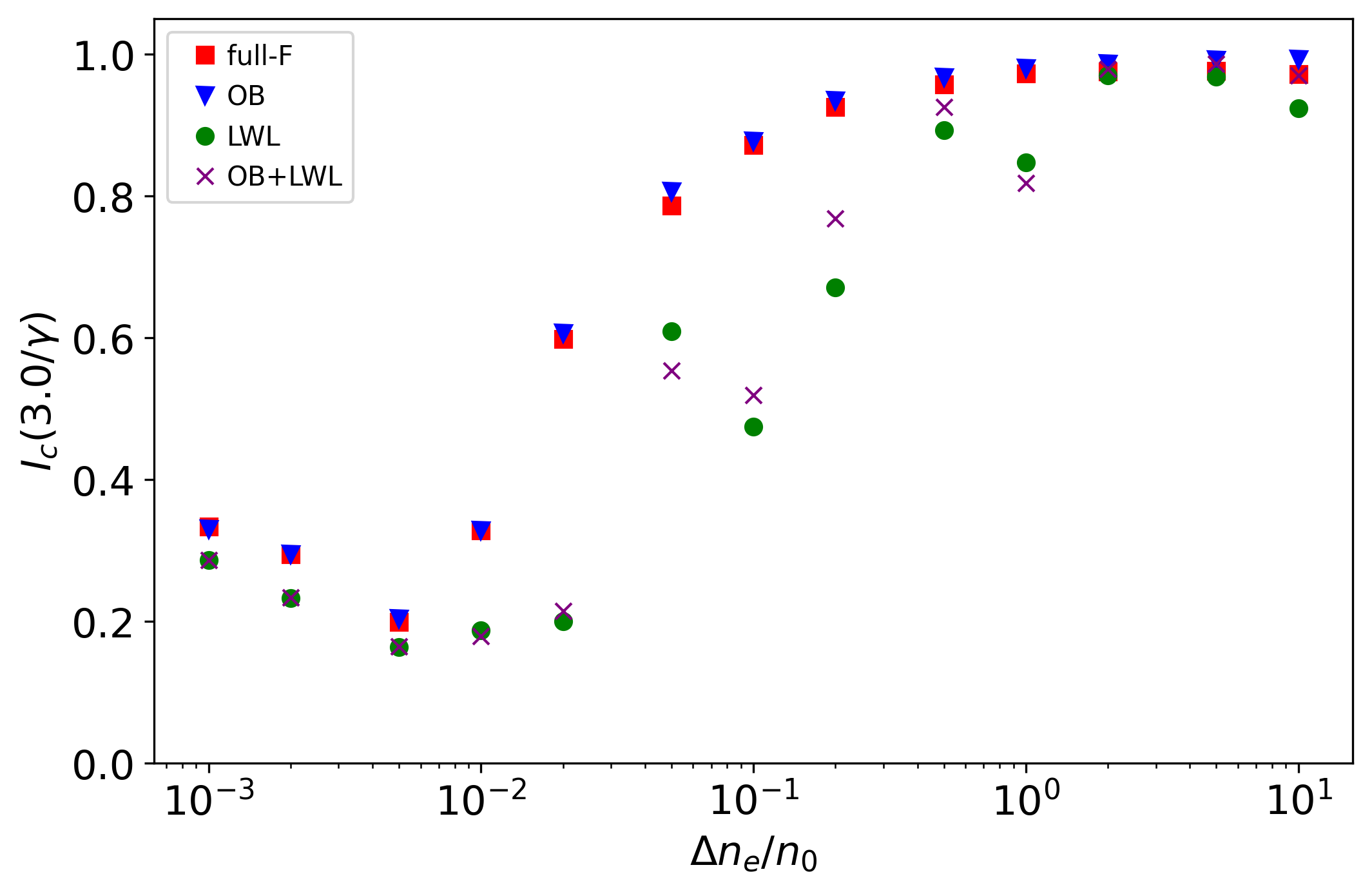}
\caption{The dependence of the blob compactness \(I_c\) at \(t=3/\gamma\) on the initial density amplitude is shown for all considered models (Table~\ref{table:poltreatment}).
}
\label{fig:IcVSamp}
\end{figure}

The blobs ability to retain its initial (Gaussian) shape is quantified by the  blob compactness
\begin{align}
     I_c(t) &:= \frac{\int dA (\ne(\vec{x},t)-\neref) h(\vec{x},t)}{\int dA 
(\ne(\vec{x},0)-\neref) h(\vec{x},0)} .
\end{align}
Here, we introduced the Heaviside function
\begin{align}
     h(\vec{x},t) &:= \begin{cases}
          1, 
        &\ \text{if} \hspace{2mm}||\vec{x} - \vec{X}_{max}||^2 < \sigma^2 \\
0,  &\ \text{else} 
           \end{cases} \nonumber, 
\end{align}
and the position of the maximum electron density \( \vec{X}_{max}(t)\).
\\
In Figure~\ref{fig:IcVStgamma} the time evolution of the blob compactness \(I_c(t)\) is shown for three different initial blob amplitudes \(\Delta \ne/\neref=\left(5.0,1.0,0.1\right)\). 
Large initial amplitude blobs decrease their compactness slower than their small initial amplitude counterparts.
Note that a slower decrease in blob compactness results in an increase in the blob-lifetime and likewise blob-coherence. 
The blob compactness is significantly reduced if the long wavelength approximation is applied, while the Oberbeck-Boussinesq approximation only slightly increases the compactness for the large initial amplitude.
\\
In Figure~\ref{fig:IcVSamp} the blob compactness \(I_c(t)\) is shown at time 
\(t=3/\gamma\) as a function of the initial blob amplitude \(\Delta \ne/\neref \) for the complete studied parameter space. 
A transition from low to high compactness takes place between roughly \(\Delta \ne/\neref \approx0.1\). Clearly, large initial blob amplitudes lead to more coherent blobs than small initial blob amplitudes. The long wavelength approximation results in significant reduction in compactness within the transition region. 
For very small blob amplitudes no significant deviations in the compactness appear if the long wavelength approximation is made, since in both cases the blobs have already largely disintegrated at \(t=3/\gamma\). 
On the other hand, for large initial blob amplitudes the disintegration of the long wavelength approximated has not been initiated at \(t=3\gamma\). Thus, in the large initial amplitude limit the long wavelength approximation introduces only a slight decrease in compactness at that time point.
The increase in coherency is attributed to the increase in \ExB vorticity when retaining arbitrary perpendicular wavelength polarization, which we introduced as gyro-amplification in Section~\ref{sec:amplification}.
\subsection{Verification of unified scaling laws for hot blobs}~\label{sec:scalinglawsverified}
\begin{figure}[ht]
\centering
\includegraphics[width=0.475\textwidth]{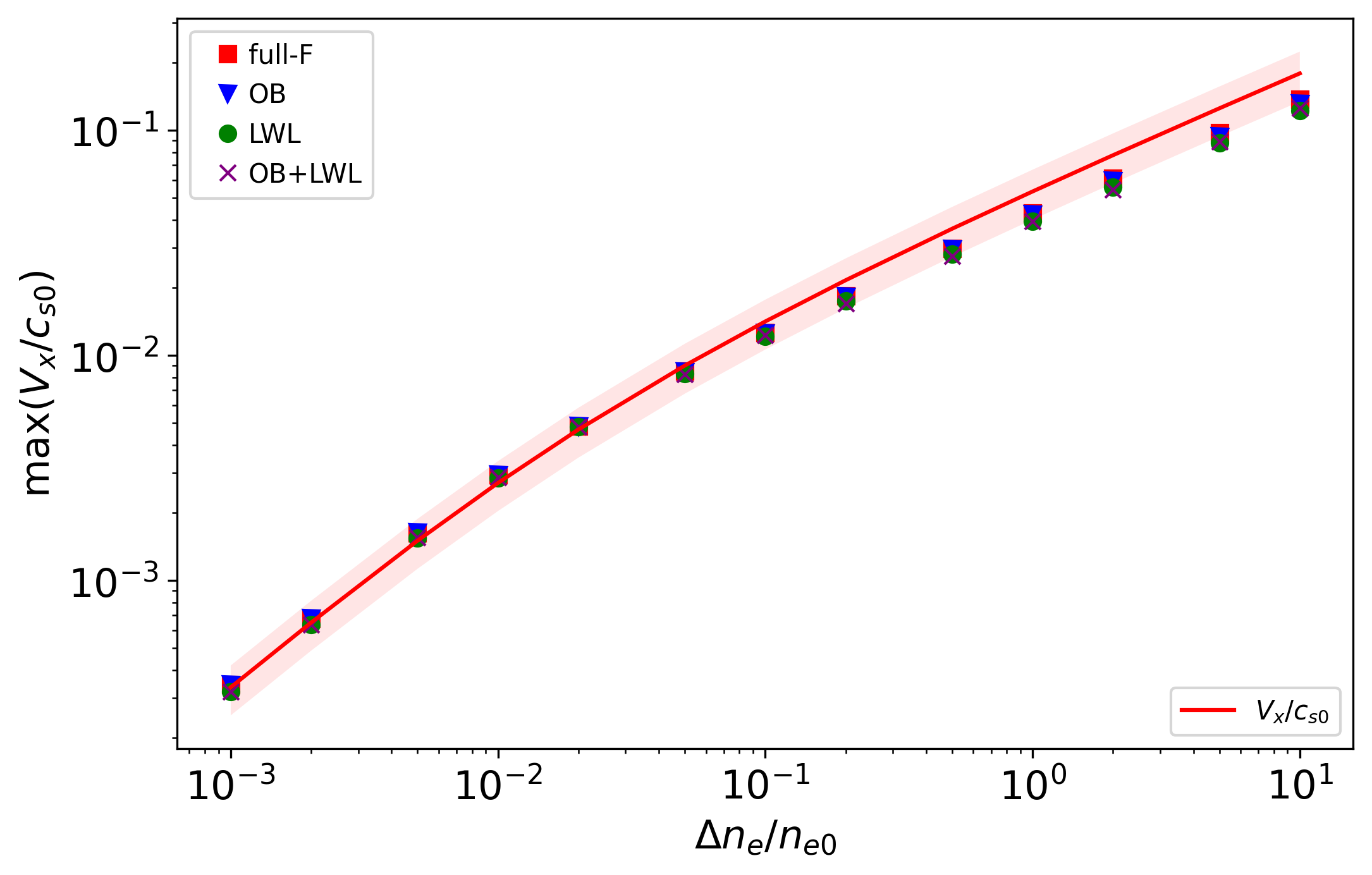}
\caption{The blob amplitude dependency of the measured maximum radial center of mass velocities are shown for all considered models (Table~\ref{table:poltreatment}). The line represents the unified velocity scaling of Eq.~\eqref{eq:Vx_unified}. The shading marks a deviation by 25\%.}
\label{fig:VxVSamp}
\end{figure}

\begin{figure}[ht]
\centering
\includegraphics[width=0.475\textwidth]{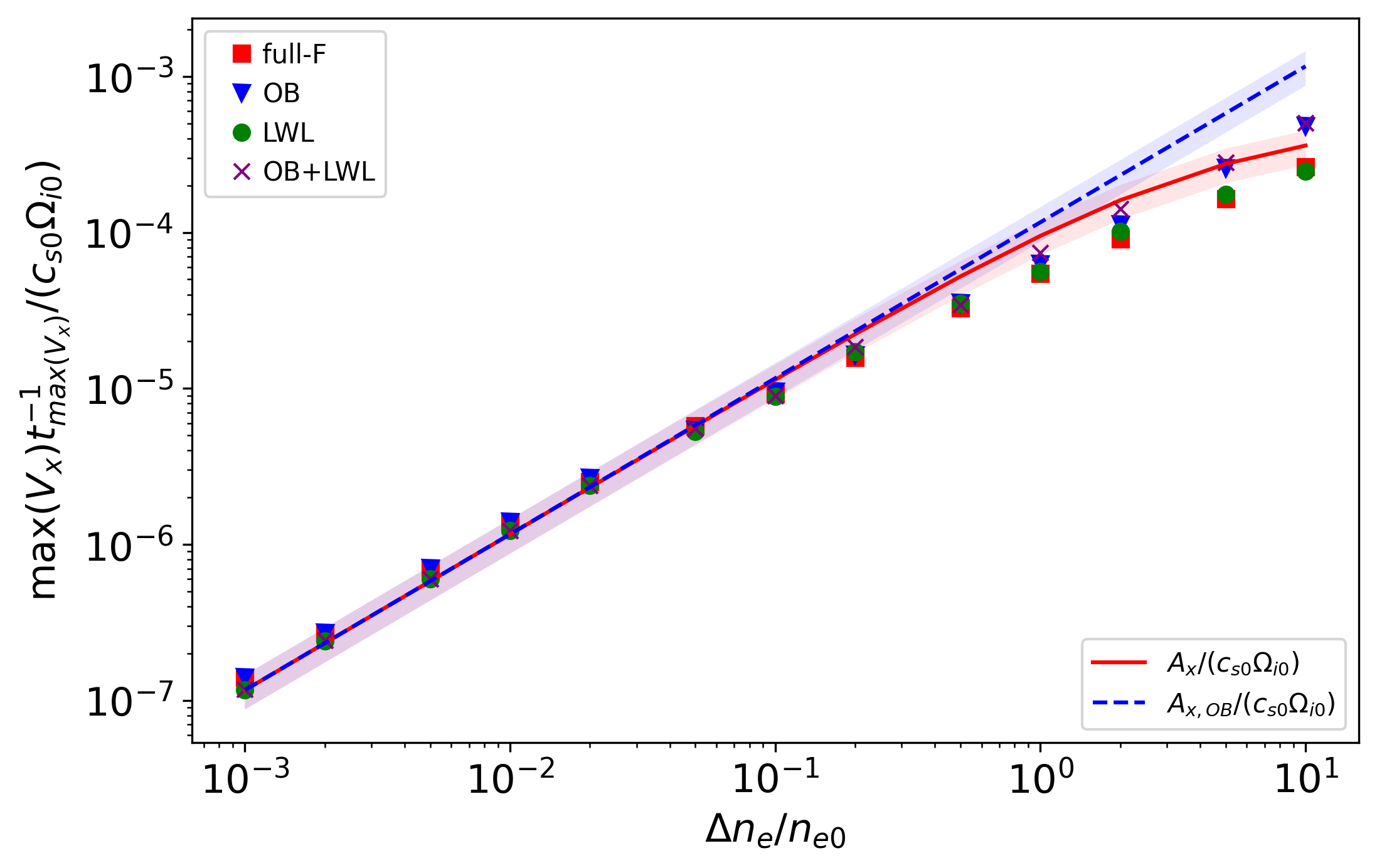}
\caption{The measured average radial center of mass acceleration as a function of the blob amplitude is depicted for all considered models (Table~\ref{table:poltreatment}) and both initial conditions. The solid and dashed line represents the acceleration scaling estimate of Eq.~\eqref{eq:acceleration} and its Oberbeck-Boussinesq limit, respectively. The shadings indicate a relative error of 25\%.}
\label{fig:AxVSamp}
\end{figure}

\begin{figure}[ht]
\centering
\includegraphics[width=0.475\textwidth]{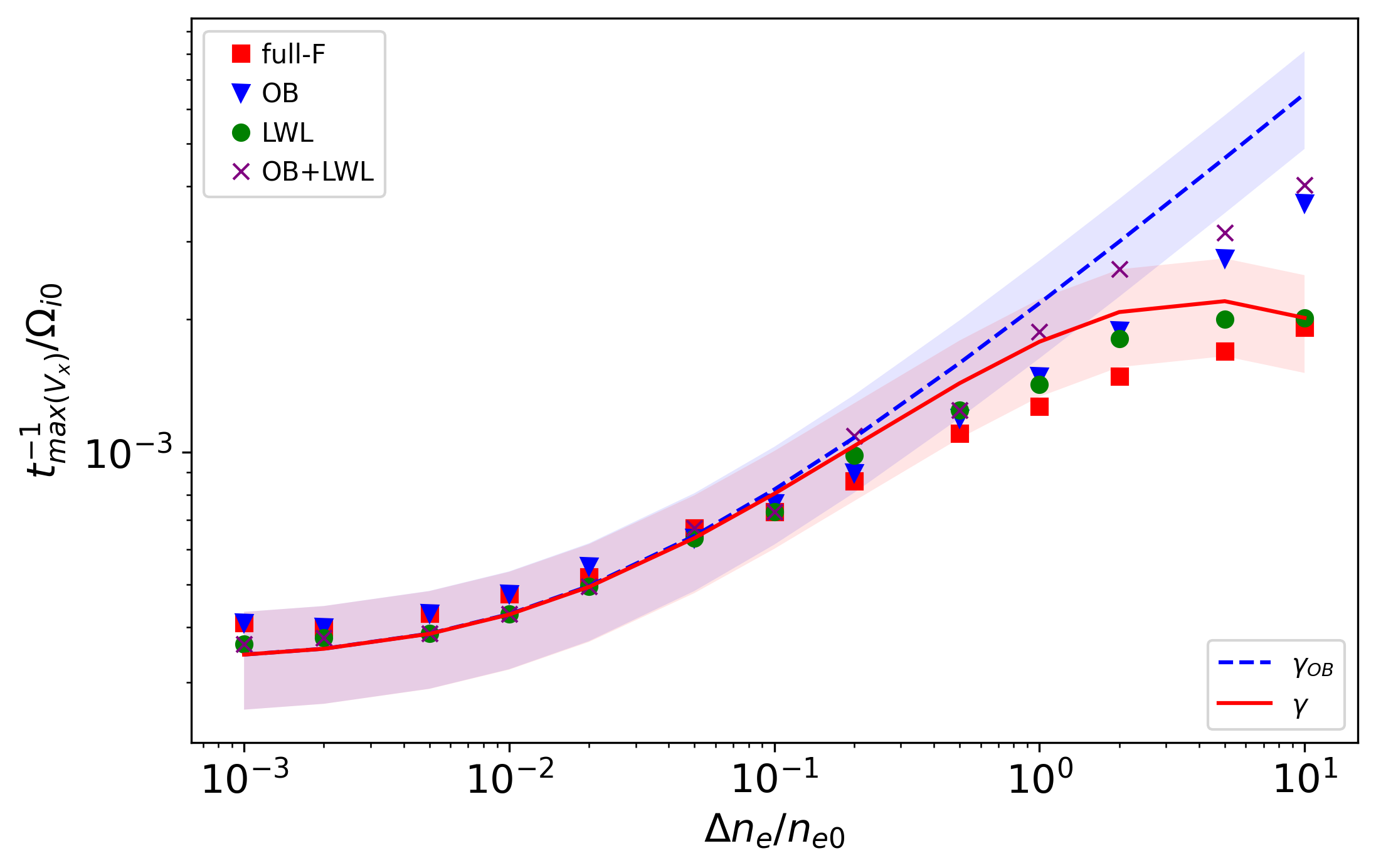}
\caption{The measured average growth rate as a function of the blob amplitude is depicted for all considered models (Table~\ref{table:poltreatment}) and both initial conditions. The solid and dashed line represents the growth rate scaling estimate of Eq.~\eqref{eq:growthrate} and its Oberbeck-Boussinesq limit, respectively. The shadings indicate a deviation by 25\%.}
\label{fig:gammavsamp}
\end{figure}

In the following we verify the derived unified scaling laws of Section~\ref{sec:scalinglaws}. We do not attempt to verify any scaling laws for the y-directed (or total) center of mass dynamics~\cite{held2016}, since this motion can depend strongly on e.g. the initial condition or on superimposed background flows of the plasma.  
\\
The fitting constants  \((\mathcal{R},\mathcal{Q})\) of the unified blob scaling laws of Eqs.~\eqref{eq:Vx_unified},~\eqref{eq:acceleration} and~\eqref{eq:growthrate} can be already determined in the cold ion limit~\cite{wiesenberger2017}. Previously, these constants have been determined by the best fit of  the acceleration and velocity scaling to \((\mathcal{R},\mathcal{Q})=(0.85,0.32)\), while neglecting the fit on the growth rate. We here present improved fitting constants  \((\mathcal{R},\mathcal{Q})=(0.95,0.31)\) determined by a best fit to all three scaling laws (velocity, acceleration and growth rate).
\\
In Figure~\ref{fig:VxVSamp} we verify the velocity scaling of Eq.~\eqref{eq:Vx_unified}  by the measured maximum of the center of mass velocities for varying initial blob amplitudes and for all considered models. 
The unified velocity of Eq.~\eqref{eq:Vx_unified} accurately captures the behaviour for all amplitudes within a relative error of 25\%. The highest accuracy is achieved for very small amplitudes. For increasing amplitudes the unified velocity scaling of Eq.~\eqref{eq:Vx_unified} slightly overestimates the center of mass blob velocities and 
 the relative error approaches roughly 25\%. 
No significant deviations due to the long wavelength or Oberbeck-Boussinesq approximation appear in the maximum of the center of mass velocities. 
\\
In Figure~\ref{fig:AxVSamp} we compare the acceleration scaling of Eq.~\eqref{eq:acceleration} with the measured linear center of mass acceleration \(\text{max} (V_x) t^{-1}_{\text{max} (V_x)}\) for varying blob amplitudes and for all considered models. We emphasize here that we do not take the absolute maximum value of the measured center of mass acceleration since our scaling theory assumes a linear acceleration (cf. Sec.~\ref{sec:scalinglaws}). In agreement with previous studies we find that the linear acceleration is increased by up to a factor two for increasing amplitudes if the Oberbeck-Boussinesq approximation is utilized. Further, the acceleration scaling estimate of Eq.~\eqref{eq:acceleration} matches the measured linear center of mass acceleration excellently for small amplitudes. For large amplitudes we find a qualitative agreement.
\\
In Figure~\ref{fig:gammavsamp} we show the growth rate scaling of Eq.~\eqref{eq:growthrate} and the measured growth rate, \(t^{-1}_{\text{max} (V_x)}\), for varying blob amplitudes and for all considered models.
The growth rate estimate of Eq.~\eqref{eq:growthrate} matches the data in the non-Oberbeck-Boussinesq regime by up to 25\%. In the Oberbeck-Boussinesq limit the agreement is slightly above the 25\% for large amplitudes, but resembles very well the respective limit of the growth rate scaling law.
\\
Note that we take always the first local maximum of the center of mass velocity and not the total maximum of the center of mass velocity, which could occur at later times for large blob amplitudes (cf.~\cite{held2016}). This is because our scaling theory assumes a linear acceleration phase that is only approximately fulfilled up to the first maximum of the center of mass velocity. However, if we take the total maximum of the center of mass velocity we obtain the same agreement of 25\% for the velocity scaling, but weaker agreement for large amplitudes for the acceleration and growth rate scaling.

\section{Discussion}\label{sec:discussion}
This paper explores for the first time the regime beyond both the long wavelength and Oberbeck-Boussinesq approximation. This previously unresolved regime is of particular importance to filamentary transport in the scrape-off layer of magnetically confined fusion devices, but also of general importance to turbulence and structure formation in magnetized plasmas.
Our study is enabled by a full-F gyro-fluid model that exploits a recently developed arbitrary wavelength polarization closure~\cite{held2020}.

Most importantly, we find that the inclusion of polarization down to the thermal gyro-radius scale leads to highly coherent blob structures in the presence of a substantial background ion temperature. The long wavelength approximation significantly reduces the coherence and lifetime of blobs due to the neglect of gyro-amplification. As a consequence, it modifies the motion of the blobs due to a faster disintegration of the blobs.

The Oberbeck-Boussinesq approximation affects the propagation of the blobs by increased linear acceleration and growth rate at large initial blob amplitudes.

The blobs center of mass motion in the linear acceleration phase is very well captured by unified velocity, acceleration and growth rate scaling laws (Eq.~\eqref{eq:Vx_unified}, ~\eqref{eq:acceleration} and~\eqref{eq:growthrate}) that we generalized to finite ion background temperature and that hold for arbitrary initial blob amplitudes.

We hypothesize that similarly to Ref.~\cite{held2016} the characteristic footprint of finite ion gyro-radius effects on the blob dynamics is amplified if a density perturbation is accompanied by a temperature perturbation.
The numerical implementation and study of an arbitrary wavelength polarization closure with more advanced full-F gyro-fluid moment hierarchies~\cite{held2020}, e.g. including temperature dynamics, parallel dynamics and electromagnetic effects, is ongoing. This effort aims to assess the validity of the long wavelength and Oberbeck-Boussinesq approximation also on turbulence in magnetized plasmas. However, the herein presented results on the permissibility of these approximation questions current efforts in edge and scrape-off layer modeling of fusion plasma that are based on at least one of this approximations.
\\
\section{Acknowledgements}
We acknowledge fruitful discussions with A. Kendl and the plasma physics and fusion energy group at Chalmers University of Technology. 
This work has been carried out within the framework of the EUROfusion Consortium, funded by the European Union via the Euratom Research and Training Programme (Grant Agreement No 101052200 — EUROfusion). Views and opinions expressed are however those of the author(s) only and do not necessarily reflect those of the European Union or the European Commission. Neither the European Union nor the European Commission can be held responsible for them. 
This work was supported by the UiT Aurora Centre Program, UiT The Arctic University of Norway (2020).
This research was funded in whole or in part by the Austrian Science Fund (FWF) [P 34241-N]. 
For the purpose of Open Access, the author has applied a CC BY public copyright license to any Author Accepted Manuscript (AAM) version arising from this submission.
  \bibliography{refs.bib}
\bibliographystyle{aipnum4-1.bst}
\end{document}